\pdfoutput=1
\documentclass[aps,pre,twocolumn]{revtex4-2}
\usepackage{amsfonts, amsmath,amssymb,float,mathtools,physics}
\usepackage{graphicx}% Include figure files
\usepackage{dcolumn}% Align table columns on decimal point
\usepackage{bm}% bold math
\usepackage[breaklinks,colorlinks = true,linkcolor = red,urlcolor=blue,citecolor=red]{hyperref}
\usepackage{multirow}
\usepackage{array}
\usepackage{booktabs}
\usepackage{ctable}
\usepackage{upgreek}
\usepackage{epsfig}
\usepackage{mathrsfs}
\usepackage{amssymb}
\usepackage{amsbsy}
\usepackage{color}
\usepackage{cancel}
\usepackage{pifont}
\usepackage{marginnote}
\usepackage{float}
\usepackage{verbatim}
\usepackage{subfigure}
\usepackage{bbm, dsfont}
\usepackage{lineno}
\usepackage{amsmath}
\definecolor{dgreen}{rgb}{0,0.7,0}

\def\bluew#1{{\color{blue} #1}}

\usepackage{natbib}
\usepackage[normalem]{ulem}

\def\bea{\begin{eqnarray}}
\def\eea{\end{eqnarray}}

\makeatletter

\newcommand{\Rmnum}[1]{\expandafter\@slowromancap\romannumeral #1@}
\makeatother

\newcommand{\beq}{\begin{equation}}
\newcommand{\eeq}{\end{equation}}

\usepackage[utf8]{inputenc}

\begin{abstract}
Resetting is a renewal mechanism in which a process is intermittently repeated after a random or fixed time. This simple act of stop and repeat profoundly influences the behaviour of a system as exemplified by the emergence of non-equilibrium properties and expedition of search processes. Herein, we explore the ramifications of stochastic resetting in the context of a single-file system called random average process (RAP) in one dimension. In particular, we focus on the dynamics of tracer particles and analytically compute the variance, equal time correlation, autocorrelation and unequal time correlation between the positions of different tracer particles. Our study unveils that resetting gives rise to rather different behaviours depending on whether the particles move symmetrically or asymmetrically. For the asymmetric case, the system for instance exhibits a long-range correlation which is not seen in absence of the resetting. Similarly, in contrast to the reset-free RAP, the variance shows distinct scalings for symmetric and asymmetric cases. While for the symmetric case, it decays (towards its steady value) as $\sim e^{-r t} / \sqrt{t}$, we find $\sim t e^{-r t}$ decay for the asymmetric case ($r$ being the resetting rate). Finally, we examine the autocorrelation and unequal time correlation in the steady state and demonstrate that they obey interesting scaling forms at late times. All our analytical results are substantiated by extensive numerical simulations.
\end{abstract}

\begin{document}

\title{Exact fluctuation and long-range correlations in a single-file model under resetting}
\author{Saikat Santra$^{1}$ and Prashant Singh$^{2}$}
\email{saikat.santrah@icts.res.in}
\email{prashant.singh@nbi.ku.dk}
\affiliation{\noindent \textit{$^{1}$ International Centre for Theoretical Sciences, Tata Institute of Fundamental Research, Bengaluru 560089, India}}

\affiliation{\noindent \textit{$^{2}$ Niels Bohr International Academy, Niels Bohr Institute,
University of Copenhagen, Blegdamsvej 17, 2100 Copenhagen, Denmark}}
\date{\today}

\maketitle

%\nocite{TitlesOn}
\section{INTRODUCTION}
Deciphering the behaviour of complex systems consisting of many interacting units is a fundamental problem often encountered in statistical physics \bluew{\cite{Lieb1966, privman_1997}}. A classic example of an interacting particle system in non-equilibrium statistical mechanics is the single-file system, in which particles in a one-dimensional line move alongside each other, strictly obeying the constraint of non-overtaking, wherein one particle cannot pass another \bluew{\cite{harris_1965, Jensen2004, 10.1214/aop/1176993602}}. Due to this non-overtaking constraint (also referred to as single-file constraint), the dynamics of different particles become strongly correlated \cite{PhysRevLett.127.220601, PhysRevE.107.044131}. For example, in a collection of diffusing particles in one dimension with single-file constraint, the mobility of a tracer particle is drastically reduced and as result, the mean-squared displacement grows sub-diffusively as $\sim \sqrt{t}$ at late times, instead of the linear growth for a freely diffusing particle \bluew{\cite{Krapivsky_2015, SadhuPRL2014, Benichou_2018, PhysRevB.18.2011}}. The coefficient of this sub-diffusive growth, in turn, depends on the particle number density, the precise interaction among the particles and also on the statistical properties of the initial state of the system \bluew{\cite{SadhuPRL2014, TSadhu2015, PhysRevE.106.L062101, PhysRevLett.102.050602, PhysRevE.88.032107}}. In fact based on hydrodynamic approach, a recent work showed that this sub-diffusive scaling holds true only for short-range interactions and changes to an interaction-dependent exponent for long-range interactions\bluew{\cite{PhysRevE.107.044129}}. Beyond diffusion, such slowing down effects have also been studied for Hamiltonian systems \cite{PhysRevLett.113.120601, ARoy2013, PhysRevLett.90.180602} as well as for other stochastic systems like randomly accelerated process \bluew{\cite{Burkhardt2019}} and active processes \cite{Teomy_2019, Galanti2013, D0SM00687D, Banerjee_2022}. In this paper, we set out to study the tracer dynamics for a single-file model in presence of a renewal mechanism called resetting that has garnered significant attention in the last decade \bluew{\cite{PhysRevLett.106.160601}}.

%Stochastic resetting is a simple and natural mechanism in which a dynamical process is intermittently interrupted after some random time, after which it again starts anew. A paradigmatic example of this is the resetting Brownian motion first studied in \bluew{\cite{..}}. In this model, a freely diffusing particle is brought back to its initial position with some rate $r$, after which it again diffuses until the next resetting event. As a result, the particle effectively experiences a confinement around its initial position. However, this confinement has a purely dynamical interpretation and does not involve physical potential. In fact, the system, at late times, reaches a non-equilibrium steady state which is characterized by non-zero probability current. Another central feature of this model is contrary to the free diffusion, the particle possesses a finite mean first-passage time which depends non-monotonically on the resetting rate $r$. This indicates that one can optimise this mean value for some optimal resetting rate $r^*$. Going beyond the standard Brownian motion, resetting has also been studied for other stochastic processes as well as in cross-disciplinary fields such as in search theory, in computer science and in chemical and biological processes. Rigorous efforts have also been made to understand non-Poissonian strategies and resetting in quantum settings.

Stochastic resetting is a simple and natural mechanism in which a dynamical process is intermittently interrupted after some random time, after which it again starts anew. A quintessential example of this phenomenon is the resetting Brownian motion, which was first studied in \bluew{\cite{PhysRevLett.106.160601}}. In this model, a particle undergoing free diffusion is returned to its starting position at a certain rate $r$, after which it recommences diffusion until the next resetting event. As a result of this, the particle experiences an effective confinement around its initial position. However, it is important to note that this confinement arises solely from the dynamics of the system and does not stem from any physical potential. Indeed, as time progresses, the system eventually reaches a non-equilibrium steady state, which is characterised by the presence of a non-zero probability current. Another notable aspect of this model is that, unlike in free diffusion, the particle exhibits a finite mean first-passage time. Remarkably, this mean time depends non-monotonically on the resetting rate $r$ which indicates its optimisation for an optimal rate $r^*$ \bluew{\cite{Evans_2011_1}}. Beyond the standard Brownian motion, resetting has also been explored within a broader spectrum of other stochastic processes \bluew{\cite{PhysRevE.91.052131, PhysRevE.91.012113, PhysRevLett.118.030603, PhysRevLett.113.220602, Bressloff_2020gfs, PhysRevLett.116.170601, Gupta_2019cjd, Singh_20202d, Evans_2018vaf, PhysRevE.102.052129, PhysRevE.103.052119,Stojkoski_2022_auto, Majumdar_2018_auto,Singh_2022aphnx, Pal_2022jafda12}} as well as in cross-disciplinary fields such as search theory \bluew{\cite{PhysRevResearch.2.043174, PhysRevLett.88.178701}}, computer science \bluew{\cite{LUBY1993173, HamlinThrasherKeyrouzMascagni+2019+329+340}} and in chemical and biological processes \bluew{\cite{doi:10.1073/pnas.1318122111, PhysRevE.93.062411, PaulBressloff_2020, PhysRevResearch.3.L032034}}. Furthermore, rigorous studies have been made to comprehend non-Poissonian strategies \bluew{\cite{PhysRevE.93.060102, PhysRevLett.118.030603, PhysRevLett.121.050601}} and the implications of resetting in quantum settings \bluew{\cite{PhysRevB.98.104309, Kulkarni2023, PhysRevLett.130.050802, Yin2023, Dubey_2023}}. On the experimental side, resetting was recently realised in experiments involving single particle in optical traps \bluew{\cite{reset-Exp1, reset-Exp2, reset-Exp3, reset-Exp3}}. We refer to \bluew{\cite{reset-review1, reset-review2, reset-review3, reset-review4, Pal_2022jafda12}} and references therein for recent reviews on the subject.

%While most of the above studies focused on non-interacting particles, there has also been a considerable interest in understanding the effect of resetting for interacting particles. Examples include exclusion processes, Ising model, fluctuating interface, predator-prey models and many others [see \bluew{\cite{..}} for a review on stochastic resetting in interacting systems]. All these studies looked at multiple particles that are simultaneously (globally) reset after some random time. This is different from some other studies where particles reset independently of the other particles.

While most of the aforementioned studies primarily focused into single particle dynamics, there has also been a substantial surge of interest in understanding the effects of resetting for interacting particles. Examples include exclusion processes \bluew{\cite{PhysRevE.100.032136, Sadekar_2020-reset, Karthika_2020-reset, Mishra_2023-reset}}, Ising model \bluew{\cite{PhysRevResearch.2.033182}}, fluctuating interfaces \bluew{\cite{PhysRevLett.112.220601, Gupta_2016_reset-FI}}, and predator-prey models \bluew{\cite{Mercado-Vsquez_2018_PP, Evans_2022-PP, PhysRevE.106.064118}}, among numerous others [see \bluew{\cite{reset-review3}} and references therein for a review on stochastic resetting in interacting systems]. These studies investigated the scenario where multiple particles are reset \textit{simultaneously} after a random duration (global resetting). This is contrary to some other studies where particles reset independently of the other particles (local resetting) \bluew{\cite{PhysRevE.106.034125, PhysRevResearch.3.L012023, Pelizzola_2021-lr}}. In a recent work involving independently diffusing particles, but undergoing global resetting at a rate $r$, the authors showed that the simultaneous resetting induces a strong long-range correlation in the system \bluew{\cite{PhysRevLett.130.207101}}. Despite this correlation, the model still possesses analytical solvability based on the renewal formula and many results on the joint probability distribution of the positions and extremal statistics were derived. However, to the best of our knowledge, the implications of stochastic resetting on the dynamics of tracer particles in an interacting multi-particle system still remains unexplored. A natural question that particularly arises - Does one still get a resetting induced long-range correlation in such interacting scenarios?  
%However, to the best of our knowledge, it still remains unknown how do inter-particle interaction and resetting affect the tracer particles in a multi-particle system. Does one still get a long-range correlation? 
In this paper, we present an example
of a single-file model (called random average process) \bluew{\cite{Schutz, PhysRevE.64.036103}} where these questions can be thoroughly addressed through exact analytic computations.

%\newpage
Our system consists of a collection of particles moving in an infinite line and distributed with density $\rho$.
%Let us first introduce our model here. We consider an infinite number of particles distributed over an infinite line with density $\rho$. 
We denote the position of $i$-th particle at time $t$ by $x_i(t)$ where $i \in \mathbb{Z}$ and $x_i(t) \in \mathbb{R}$. Initially, the particles are located at a fixed distance $a = \frac{1}{\rho}$ apart:
\begin{align}
x_i(0) = i a = i/ \rho,~~~~~\text{for all }i \in \mathbb{Z}.
\label{ini-pos}
\end{align}
For simplicity, we take $\rho=1$ without any loss of generality.
%Let us now discuss the time evolution of the positions of the particles. 
Starting from this configuration, each particle performs the random average process interspersed by resetting events, during which the entire system is (globally) reset to the configuration in Eq.~\eqref{ini-pos}. This means that at any small time interval $[t, t+\Delta  t]$, two events can occur: (i) either particles globally reset their positions to $x_i(0)$ in Eq. \eqref{ini-pos} with probability $r \Delta t$ (ii) or they perform the random average motion with the remaining probability $(1-r\Delta t)$. In the case of latter event, the $i$-th particle can jump to its right with probability $p \Delta t $ and to its left with probability $q \Delta t$. With remaining probability $\left[1- (p+q) \Delta t\right]$, the position $x_i(t)$ does not change. The successful jump, either to the left or to the right, is by a random fraction $\eta _i$ of the space available between the particle and its neighbour. This means that the jump to the right takes place by an amount $\eta _i \left[ x_{i+1}(t) - x_i(t)\right]$ while to the left, the particle jumps by the amount $\eta _i \left[ x_{i-1}(t) - x_i(t)\right]$. Here $\eta \in  [0,1)$ is a random variable drawn from the distribution $R(\eta)$. Notice that throughout time evolution, a particle can never overtake its neighbouring particles and maintains its initial order.

The overall update rule for the position can then be written as 
\begin{align}
x_i(t+ \Delta t) =
\begin{cases} & x_i(0), ~~~~~~~~~~~~~\text{w.p.   } r \Delta t,\\
& x_i (t) + \Gamma _i (t),~~~~ \text{w.p.  } (1-r \Delta t),
\end{cases}
\label{update-rule}
\end{align}
where `w.p.' stands for `with probability' and $\Gamma _i (t)$ for the increment which is given by
\begin{align}
\Gamma _i (t) = 
\begin{cases}
& \eta _i \left[ x_{i+1}(t) - x_i(t)\right],~~\text{w.p. }p \Delta t, \\
& \eta _i \left[ x_{i-1}(t) - x_i(t)\right],~~\text{w.p. }q \Delta t, \\
& ~~~~0, ~~~~~~~~~~~~~~~~~~~~~\text{w.p. }\left[1-(p+q) \Delta t \right].
\end{cases}
\label{update-rule-2}
\end{align}
Given this model, our aim in this paper is two fold: (i) Firstly, we aim to study the dynamics of tracer particles and explore analytically the effect of stochastic resetting on the single-file model. For this purpose, we investigate the variance and different two-point correlation functions for the positions of tracer particles. In particular, the role of resetting on the autocorrelation involving two different times has so far been studied only for few cases such as the drift-diffusion \bluew{\cite{Stojkoski_2022_auto}} and the fractional Brownian motion \bluew{\cite{Majumdar_2018_auto}}. It also played a crucial role in income dynamics modelled using geometric Brownian motion \cite{income-Pal}. However, all of these studies dealt with the single particle dynamics. Here, our aim is to investigate this with multiple particles and understand how fluctuations corresponding to one particle at a given time affect the position of some other particle at some later time. (ii) Secondly, our work will also shed light on determining whether the long-range correlation, stemming from the simultaneous resetting of non-interacting particles \cite{PhysRevLett.130.207101}, persists even when the system involves interacting particles in its underlying dynamics. 
%Secondly, this model presents one of the few examples where correlation functions involving two different times can be computed exactly. %To the best of our knowledge, this calculation has so far been possible only for fractional Brownian motion and 

%Firstly, it will answer whether the long range correlation induced by simultaneous resetting of otherwise non-interacting particles remain present even when the underlying dynamics has interacting particles.

%Our aim is to study the (local) dynamics of the tracer particles exploring the ramifications of non-zero $r$ on them. Due to resetting, we anticipate the tracer particles to attain a non-equilibrium steady state at long times. But the question is - how does the system relax to this steady state? Are the relaxation properties same for the unbiased $(p=q)$ and the biased $(p \neq q)$ cases? How do the positions of various tracer particles get correlated in the steady state? This paper will present a systematic analytical answers to these questions for the model in Eq.~\eqref{update-rule}.

The remainder of our paper is organized as follows: Section \ref{sec-no-reset} briefly recalls the main results of the reset-free random average process. In Section \ref{sec-avg-pos}, we illustrate the effect of resetting on the mean position of a tracer particle. These results will then be used to calculate the variance in Section \ref{sec-var} and equal time correlation in Section \ref{sec-corr}. We devote Sections \ref{sec-auto} and \ref{sec-uneq} to compute respectively, the autocorrelation and the unequal time correlation in the steady state. Finally, we conclude in Section \ref{sec-conclusion}

%We first briefly recall the main results of the standard random average process in Section \ref{sec-no-reset}.
%Section \ref{sec-no-reset}
%Section \ref{sec-avg-pos}
%Section \ref{sec-var}
%Section \ref{sec-corr}
%Section \ref{sec-auto}
%Section \ref{sec-uneq}

\section{Random average process without resetting}
\label{sec-no-reset}
Before we delve into discussing the consequences of resetting on tracer particles, it is instructive to review some known results for the random average process (henceforth RAP) in absence of resetting $(r=0)$. Studied originally by Fontes and Ferrari as a generalisation of the smoothing process and the voter model \bluew{\cite{10.1214/EJP.v3-28}}, the RAP has appeared in several physical problems like force fluctuations in bead packs \bluew{\cite{PhysRevE.53.4673}}, in mass transport models \bluew{\cite{Rajesh2000con, Krug000con}}, in models of wealth distribution and traffic \bluew{\cite{Ispolatov1998}}
and generalized Hammersley process \bluew{\cite{Hammersely1995}}. As an interacting multi-particle system, this model is particularly interesting since the particles do not overtake each other and maintain their initial order throughout the time evolution as indicated by Eq.~\eqref{update-rule}. Therefore, every particle performs the single-file motion. Due to this non-overtaking constraint, the motion of different particles get strongly correlated \cite{Schutz, PhysRevE.64.036103, Cividini_2016_RAP, Kundu_2016_RAP, Cividini_2016_RAPP, Cividini_2017_RAP}. For instance, various studies based on both microscopic calculations as well as hydrodynamic approaches have shown that the variance and the correlation of the displacement variable $z_i(t)$ (with $z_i(t) = x_i(t)-x_i(0)$) at late time is given by \bluew{\cite{PhysRevE.64.036103}}
%Study on the motion of tagged particles for RAP has been extensively carried out in \bluew{\cite{..}}. These studies reveal that the fluctuations and correlations of their positions exhibit distinct scaling behaviours \bluew{\cite{..}} characterised by non-trivial scaling functions. Moreover, these scaling behaviours are found to depend quite sensitively on the initial preparation of the system. For instance, in the quenched case (where the initial positions are fixed to Eq. \eqref{ini-pos} for every realisation), the mean, variance and the connected-correlation of the displacement variable $z_i(t) = x_i(t)-x_i(0)$ scales as
\begin{align}
 & \langle z_i(t) \rangle = \mu _1(p-q) t,  \label{mean-no}\\
&  \langle z_i^2(t) \rangle-\langle z_i(t) \rangle^2 \simeq   \zeta \sqrt{t}, \label{msd-no} \\
& \langle z_0(t) z_i(t) \rangle-\langle z_i(t) \rangle \langle z_0(t) \rangle \nonumber \\
& ~~~~~~~~~~~~~~~\simeq \zeta \sqrt{t}~f \left( \frac{|i|}{2 \sqrt{\mu _1(p+q)t}} \right), \label{corr-no}
\end{align}
where the pre-factor  $\zeta = \mu _2 \sqrt{\mu _1(p+q)} / \left[ \sqrt{\pi}(\mu _1-\mu _2) \right]$ depends on first two moments $\mu _1$ and $\mu _2$ of the jump distribution $R(\eta)$ and the scaling function $f(y)$ in Eq. \eqref{corr-no} is given by
\begin{align}
f(y) = e^{-y^2} - \sqrt{\pi} y~ \text{Erfc}(y). \label{scaling-no}
\end{align}
Due to the translational symmetry in the model, both mean and variance in Eqs.~\eqref{mean-no} and \eqref{msd-no} do not depend on the particle index $i$. Also the mean vanishes for the symmetric case $p=q$ since particles do not experience any drive under this condition. Meanwhile the sub-diffusive scaling of the variance at large times in Eq. \eqref{msd-no} and the scaling function $f(y)$ in Eq. \eqref{scaling-no} are hallmark properties of many single-file systems that possess diffusive hydrodynamics at the macroscopic scales \bluew{\cite{Cividini_2017_RAP, SadhuPRL2014}}.

%is hallmark of many single-file systems although the pre-factor $\zeta$ depends on the specific choice of model \bluew{\cite{..}}.
%Few remarks are in order: First, the mean in Eq. \eqref{mean-no} vanishes in the symmetric case $p=q$ since the particle does not experience any drive under this condition. Also, due to the translational symmetry, both these quantities are independent of the index $i$. Finally, we remark that the sub-diffusive scaling of the variance at large times in Eq. \eqref{msd-no} is a hallmark of many single-file systems although the pre-factor $\zeta$ depends on the specific choice of model \bluew{\cite{..}}.

It turns out that the variance and the correlation for single-file systems depend crucially on the statistical properties of the initial state of the system. For RAP, results in Eqs.~\eqref{mean-no}-\eqref{scaling-no} hold true only for the quenched initial condition where the initial positions are fixed to Eq.~\eqref{ini-pos} for all realisations \bluew{\cite{PhysRevE.64.036103}}. On the other hand, for annealed initial condition where initial positions are drawn from the steady-state, the initial positions themselves vary from realisation to realisation. Under this circumstance, the variance $l_0(t)$ at late time grows as 
%So far, we have discussed results only for the quenched (fixed) initial condition. However, in the annealed case, the initial positions can fluctuate from realisation to realisation, and one then has to perform an additional averaging over them. In particular, if we measure the displacement variable in the steady state as $r_i(t) = \lim _{t_0 \to \infty} \left[ x_i(t_0+t)-x_i(t_0) \right]$, then the variance $l_0(t) = \langle r_i^2(t) \rangle - \langle r_i(t) \rangle ^2$ in the annealed case scales at late times as
\begin{align}
l_0(t) & \simeq \zeta \sqrt{2 t},~~~~~~~~~~~~~~~\text{for }p=q, \\
& \simeq \frac{\mu _1 \mu _2 (p-q)}{(\mu _1 - \mu _2)} t, ~~~~~\text{for }p>q.
\end{align}
As indicated, the temporal scaling of $l_0(t)$ depends sensitively on whether particles experience drive or not. This is contrary to the quenched case in Eq.~\eqref{msd-no} where we obtain same scaling for both cases. Furthermore, in the case of symmetric RAP, the ratio of the variances for two cases is found to be $\sqrt{2}$ which is also observed in the context of other single-file models \bluew{\cite{SadhuPRL2014, PhysRevE.106.L062101}}. In what follows, we will investigate these quantities for non-zero $r$ and illustrate how resetting modifies them.

%Interestingly, unlike in the quenched case, the asymptotic behaviour of the variance depends on whether the jump is symmetric $(p=q)$ or antisymmetric $(p \neq q)$. While in the asymmetric case, the variance at large times scales linearly with time, we still recover the sub-diffusive growth for the symmetric case (reminiscent of the quenched case). But the pre-factor associated with this sub-diffusive growth differs by a factor of $\sqrt{2}$ compared to the quenched case in Eq. \eqref{msd-no}. This relation between the variances at two different initial conditions has also been observed in other single-file systems \bluew{\cite{..}}. In the following, we study the effect of resetting on the variance and correlation functions and analyse their dependence on the initial history of the system. 

\section{Average position}
\label{sec-avg-pos}
\begin{figure}[t]
\includegraphics[scale=0.8]{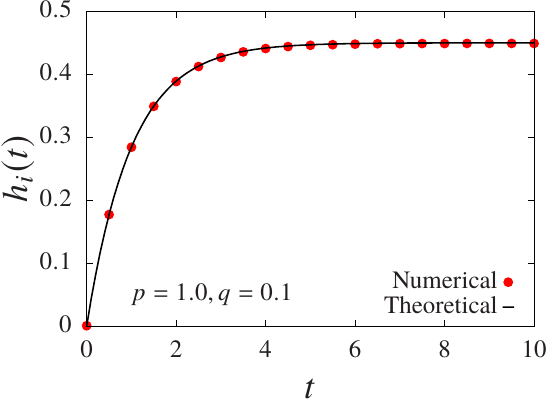}
\centering
\caption{Comparison of the mean $h_i(t) = \langle z_i(t) \rangle$ in Eq. \eqref{avg-eq-3} with the numerical simulation for $p=1.0,~q=0.1,~r=1$ and jump distribution $R(\eta)=1$. Simulation is conducted with $N=101$ particles.}
%two sets of parameters : $p=1.0,q=0.1$ in panel (a) and $p=0.5,q=0.25$ in panel (b). The red circular points represent numerical result whereas the black solid line corresponds to the theoretical expression. For both panels, we have chosen $N=101,r=1$. \redw{\textbf{Saikat, please put this plot only for one value of $p$ and $q$ and for $R(\eta)=1$}}}    
\label{avg-pic-1}
\end{figure}

For $p \neq q$, we saw in Eq.~\eqref{mean-no} that the particles experience a net drive which gives rise to a non-zero value of mean that grows linearly with time. Let us investigate what happens to this average in presence of resetting. Here again, it turns out convenient to work in terms of the displacement variable $z_i(t) = x_i(t)-x_i(0)$ and rewrite the update rules in Eq.~\eqref{update-rule} as
%We first study the effect of resetting on the average of the displacement variable $z_i(t) = x_i(t)-x_i(0)$. Rewriting update rule in Eq. \eqref{update-rule} in terms of $z_i(t)$, we get
\begin{widetext}
\begin{align}
z_i(t+ \Delta t) =
\begin{cases} & 0, ~~~~~~~~~~~~~~~~~~~~~\text{with probability } r \Delta t,\\
& z_i (t) + \Gamma _i (t),~~~~~~~ \text{with probability } (1-r \Delta t),
\end{cases}
\label{update-rule-3}
\end{align}
where the increment $\Gamma _i (t)$ can be written as
\begin{align}
\Gamma _i (t) = 
\begin{cases}
& \eta _i \left[ z_{i+1}(t) - z_i(t)+1 \right],~~~\text{with probability }p \Delta t, \\
& \eta _i \left[ z_{i-1}(t) - z_i(t) -1\right],~~~\text{with probability }q \Delta t, \\
& 0, ~~~~~~~~~~~~~~~~~~~~~~~~~~~~~~\text{with probability }1-(p+q) \Delta t.
\end{cases}
\label{update-rule-4}
\end{align}
\end{widetext}
Denoting the mean as $h_i(t) = \langle z_i(t) \rangle $, we can write its evolution for a small time interval $[t,t+\Delta t]$ as
%Here, we are interested in calculating the mean $h_i(t) = \langle z_i(t) \rangle $ where average $\langle ...  \rangle$ is considered over all histories. For this, we consider $h_i(t +\Delta t) = \langle z_i(t+\Delta t) \rangle $ and then insert Eqs. \eqref{update-rule-3} and \eqref{update-rule-4} to obtain
\begin{align}
h_i(t +\Delta t) & \simeq (1-r \Delta t) \left[ h_i(t) + \langle \Gamma _i(t) \rangle  \right].
\end{align}
Plugging $\langle \Gamma _i(t) \rangle$ from Eq. \eqref{update-rule-4} and taking $\Delta t \to 0$ limit, we obtain the following differential equation for $h_i(t)$:
\begin{align}
\frac{d h_i(t)}{dt} &= \mu _1  \left[ p h_{i+1}(t) + q h_{i-1}(t)-(p+q) h_{i}(t) \right]  \nonumber \\
&~~~~~~~~  - r h_i(t) +\mu _1(p-q).
\label{avg-eq-1}
\end{align}
Recall that $\mu _m = \langle \eta ^m \rangle = \int _0 ^{1} d\eta ~\eta ^m R(\eta) $ represents the $m$-th moment of the jump distribution $R(\eta)$. One needs to solve this equation with the initial condition $h_i(0) = 0$. Since it is a linear equation in $h_i(t)$, we proceed to solve it by taking the Fourier transformation with respect to the index $i$. Defining the Fourier transformation as
\begin{align}
\bar{h}(k,t) = \sum _{i=-\infty}^{\infty}  h_i(t)~e^{j i k},
\label{FT}
\end{align}
and the inverse Fourier transform as
\begin{align}
h_i(t) = \frac{1}{2 \pi} \int _{-\pi}^{\pi} dk ~\bar{h}(k,t)~e^{-j i k},
\label{IFT}
\end{align}
with $j^2 = -1$, we recast Eq.~\eqref{avg-eq-1} in terms of the Fourier variable as
\begin{align}
\frac{\partial \bar{h}(k,t)}{\partial t}  = -\alpha (k)~\bar{h}(k,t)+ 2 \pi \mu _1 (p-q) \delta (k),
\label{avg-eq-2}
\end{align}
with $\alpha (k) = r+\mu _1(p+q)-\mu_1 p e^{j k}-\mu_1 q e^{-j k}$. Finally solving Eq.~\eqref{avg-eq-2} and performing the inverse Fourier transformation, we get
\begin{align}
h_i(t) = h(t) =\frac{\mu _1 (p-q)}{r} \left( 1- e^{- r t}\right).
\label{avg-eq-3}
\end{align}
Notice that the final expression turns out to be independent of the index $i$, since we have assumed translational invariance of our infinite system. Also, the mean expectedly vanishes for the symmetric $p=q$ case. Meanwhile for $r = 0$, our result reduces to Eq. \eqref{mean-no} where mean grows linearly with time. However, for any non-zero $r$, it approaches a steady value at large times. This steady value decays as $\sim 1 /r$ with respect to the resetting rate. Physically, a larger resetting rate confines the tracer particle to move in the vicinity of its initial position which gives rise to the smaller mean. In Figure \ref{avg-pic-1}, we have plotted the mean $h_i(t)$ and compared it with the same obtained from numerical simulations. We observe an excellent match between them. In the following sections, we will use this expression of mean to compute different correlation functions for the tracer particles.

%We have performed the comparison for two different choices of the jump distribution, namely, $R(\eta)=1$ and $R(\eta) = 3(1-\eta)^2$. An excellent match is observed for both choices.

%\begin{figure*}[t]
 %\includegraphics[width=\textwidth]{figure1.pdf}
  %\centering
%  \subfigure{\includegraphics[scale=0.35]{avg-uniform.pdf}}
 % \subfigure{\includegraphics[scale=0.35]{avg-pos-power.pdf}}
%\centering
%\caption{Comparison of the mean $h_i(t) = \langle z_i(t) \rangle$ in Eq. \eqref{avg-eq-3} with the numerical simulation for two sets of parameters : $p=1.0,q=0.1$ in panel (a) and $p=0.5,q=0.25$ in panel (b). The red circular points represent numerical result whereas the black solid line corresponds to the theoretical expression. For both panels, we have chosen $N=101,r=1$. \redw{\textbf{Saikat, please put this plot only for one value of $p$ and $q$ and for $R(\eta)=1$}}}    
%\label{avg-pic-1}
%\end{figure*} 
\begin{figure*}[t]
	
	\includegraphics[width=\textwidth]{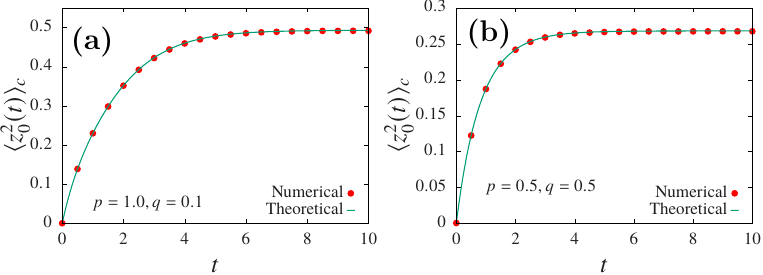}
	%\centering
	%  \subfigure{\includegraphics[scale=0.35]{avg-uniform.pdf}}
	% \subfigure{\includegraphics[scale=0.35]{avg-pos-power.pdf}}
	%\centering
	\caption{Comparison of $ \langle z_0^2(t) \rangle_c$ in Eq. \eqref{eq-corr-eq-11} with the numerical simulation for two sets of parameters : $p=1.0,~q=0.1$ in panel (a) and $p=0.5,~q=0.5$ in panel (b). For both panels, we have performed simulation with $N=101$ particles, resetting rate $r=1$ and jump distribution $R (\eta)=1$.} 
%The red circular points represent numerical result whereas the black solid lines correspond to their theoretical expressions.    
	\label{msd-pic-1}
\end{figure*} 
\section{Variance and Equal time correlation}
In this section, we look at the variance and equal time correlations of the positions of two tagged particles when the entire system is reset to the configuration in Eq.~\eqref{ini-pos} with rate $r$. Let us denote this correlation by $C_i(t)=\langle z_{0}(t) z_{i} (t) \rangle $. Recall that for free RAP, the correlation function satisfies a scaling behaviour in $\big|i \big| /\sqrt{t}$ with the associated scaling function given in Eq.~\eqref{scaling-no}. Here, our aim is to illustrate how this scaling behaviour gets modified in presence of resetting. To this aim, we start by deriving the time evolution differential equation for $C_i(t)$. In a small time interval $[t, t+\Delta t]$, the correlation $C_i \left( t + \Delta t \right) $ for $i \neq 0$ changes by an amount
\begin{align}
C_i \left( t + \Delta t \right) & \simeq C_i \left( t \right)-r \Delta t ~ C_i \left( t \right)+2 \mu _1 (p-q) \Delta t  ~h(t)
\nonumber \\
& +\mu _1 (p+q) \Delta t \left[ C_{i+1}(t)+C_{i-1}(t)-2 C_{i}(t) \right]. \nonumber 
\end{align}
On the other hand, applying the same procedure for $i =0$ gives
\begin{align}
C_0 \left( t + \Delta t \right)  & \simeq  C_0\left( t \right)-r \Delta t ~ C_0 \left( t \right)+2 \mu _1 (p-q) \Delta t  ~h(t)
\nonumber \\
&  +\mu _2(p+q)\Delta t ~\left[ 1+2 \{ C_0(t)-C_1(t)\} \right] \nonumber \\
& +\mu _1 (p+q) \Delta t \left[ 2C_{1}(t)-2 C_{i}(t) \right] \nonumber
\end{align}
%, at late times, satisfies distinct scaling relation as seen in Eq. \eqref{corr-no}.
%with resetting positions fixed to Eq. \eqref{ini-pos}. Recall that for free RAP, the correlation function, at late times, satisfies distinct scaling relation as seen in Eq. \eqref{corr-no}. Here, our aim is to illustrate how this relation gets modified in presence of resetting. Introducing the notation $C_i(t)=\langle z_{0}(t) z_{i} (t) \rangle $, we again consider its evolution in a small time interval $[t, t+\Delta t]$ and plug $z_0(t+\Delta t)$ and $z_i(t+\Delta t)$ from Eq. \eqref{update-rule-3}. Retaining terms up to linear order in $dt$ and then taking $dt \to 0$ limit, we obtain 
Combining both contributions for $i=0$ and $i \neq 0$ and taking the limit $\Delta t \to 0$, we obtain
\begin{align}
\frac{d C_i(t)}{d t} =&  \mu _1(p+q) \left[ C_{i+1}(t)+C_{i-1}(t)-2 C_{i}(t) \right] \nonumber \\
& +\delta _{i,0}~ \mu _2(p+q) \left[ 1+2 \{ C_0(t)-C_1(t)\} \right] \nonumber \\
&~~~~~ + 2 \mu _1 (p-q) h(t)-r C_{i}(t).
\label{eq-corr-eq-1}
\end{align}
Fortunately, this equation involves only mean and two-point correlation function and does not involve higher order correlation functions. This closure property enables us to obtain exact solution for this equation. For this, let us take the Laplace transformation with respect to $t~(\to s)$ as
\begin{align}
\hat{C}_i(s) = \int _0 ^{\infty} dt ~e^{-s t} ~C_i(t),
\label{LT}
\end{align}
and rewrite Eq.~\eqref{eq-corr-eq-1} in terms of $\hat{C}_i(s)$ as
\begin{align}
s \hat{C}_i(s)=&~~  \mu _1(p+q) \left[ \hat{C}_{i+1}(s)+\hat{C}_{i-1}(s)-2 \hat{C}_{i}(s) \right] \nonumber \\
&+\delta _{i,0} ~\mu _2(p+q) \left[ \frac{1}{s}+2 \{ \hat{C}_{0}(s)-\hat{C}_{1}(s)\} \right] \nonumber \\
& ~~~~~~~ -r \hat{C}_{i}(s) + 2 \mu _1 (p-q) \hat{h}(s).
\label{eq-corr-eq-2}
\end{align}
While writing this equation, we have taken the initial condition $C_i(0)=0$ and introduced the notation $\hat{h}(s)$ to denote the Laplace transformation of the mean $h(t)$ which from Eq. \eqref{avg-eq-3} turns out to be
\begin{align}
\hat{h}(s) = \frac{\mu _1 (p-q)}{r} \left(  \frac{1}{s}-\frac{1}{s+r} \right). \label{ashbjvie}
\end{align}
We now proceed to solve Eq. \eqref{eq-corr-eq-2}. First notice that, due to the single-file constraint, we get coupling of different $i$ in Eq. \eqref{eq-corr-eq-2}. To decouple them, we take the Fourier transformation
\begin{align}
\mathcal{Z}(k,s) = \sum _{i=-\infty}^{\infty} \hat{C}_i(s) ~e^{jik},
\end{align}
and insert this in Eq. \eqref{eq-corr-eq-2} to yield
\begin{align}
\mathcal{Z}(k,s) = &  \frac{\mu _2(p+q) \left[1 +2 s \{ \hat{C}_0(s)-\hat{C}_1(s) \} \right]}{s \left[ s+r+2 \mu _1 (p+q)(1-\cos k)\right]} \nonumber \\
&~~~~~~~~ + \frac{4 \pi \mu _1 (p-q)\hat{h}(s)}{(s+r)} ~\delta (k). 
\label{eq-corr-eq-3}
\end{align}
Everything on the right hand side is known except two functions, namely $\hat{C}_0(s)$ and $\hat{C}_1(s)$. One of them can be expressed in terms of the other by putting $i=0$ in Eq. \eqref{eq-corr-eq-2}. This results in the relation
\begin{align}
\hat{C}_1(s) & = \frac{\left[ 2(p+q)(\mu_1-\mu _2) +s+r\right] \hat{C}_0(s) }{2 (\mu _1 - \mu _2)(p+q)} \nonumber \\
& ~~~~   - \frac{\mu _2(p+q)/s + 2 \mu _1(p-q) \hat{h}(s)}{ 2 (\mu _1 - \mu _2)(p+q) },
\label{eq-corr-eq-4}
\end{align}
which we substitute in Eq. \eqref{eq-corr-eq-3} to yield
\begin{align}
\mathcal{Z}(k,s) = &  \frac{\mu _2}{s} \frac{ \mu _1(p+q)-s (s+r) \hat{C}_0(s) + 2 \mu _1 s (p-q) \hat{h}(s)  }{(\mu _1 - \mu _2)\left[ s+r+2 \mu _1(p+q)(1-\cos k) \right]}         \nonumber \\
&~~~~~~~~ + \frac{4 \pi \mu _1 (p-q)\hat{h}(s)}{(s+r)} ~\delta (k). 
\label{eq-corr-eq-343}
\end{align}
We now have only one unknown $\hat{C}_0(s)$. However, this can be computed self-consistently as shown later. Finally inverting $\mathcal{Z}(k,s) $ in $k$ gives the exact correlation as
\begin{align}
& \hat{C}_i(s)  = \frac{2 \mu _1 (p-q)\hat{h}(s)}{s+r}  + \hat{\mathcal{G}}_i(s), \label{eq-corr-eq-5}
\end{align}
where the function $\hat{\mathcal{G}}_i(s)$ is defined as
\footnotesize{\begin{align}
\hat{\mathcal{G}}_i(s) & = \frac{ \mu _2 \left[ \mu _1(p+q)-s (s+r) ~\hat{\mathcal{G}}_0(s)\right]}{s(\mu _1 - \mu _2)} \times  \mathcal{V}_i(s+r), \label{eq-corr-eq-6}\\
\mathcal{V}_i(s) &= \frac{1}{2 \pi } \int _{-\pi}^{\pi} dk~\frac{e^{-j i k}}{s + 2 \mu _1' (1-\cos k)}, \nonumber \\
& = \frac{1}{\sqrt{s^2 + 4 \mu _1 ' s}} \left[ \frac{s +2 \mu _1 ' -\sqrt{s^2 + 4 \mu _1' s} }{2 \mu _1' }\right]^{|i|}
\end{align}}
\normalsize{with} $\mu _1 ' = \mu _1(p+q)$. To summarise, we have calculated the exact two-point correlation functions in Eq. \eqref{eq-corr-eq-5} in terms of the Laplace variable $s$. The idea now is to perform the inversion and obtain them in the time domain. For simplicity, we carry out this inversion separately for $i=0$ and $i \neq 0$ cases below.

%\begin{figure}[t]
%\includegraphics[scale=0.3]{msd-uniform.pdf}
%\includegraphics[scale=0.3]{msd-power.pdf}
%\centering
%\caption{Comparison of $ \langle z_0^2(t) \rangle$ in Eq. \eqref{eq-corr-eq-11} with the numerical simulation for jump distribution $R(\eta) = 1$ [panel (a)] and $R(\eta) = 3(1-\eta)^2$ [panel (b)]. For both panels, we have chosen $p=0.5,~q=0.25$ and $r=1$.}    
%\label{msd-pic-1}
%\end{figure}

\begin{figure*}[t]
	
	\includegraphics[width=\textwidth]{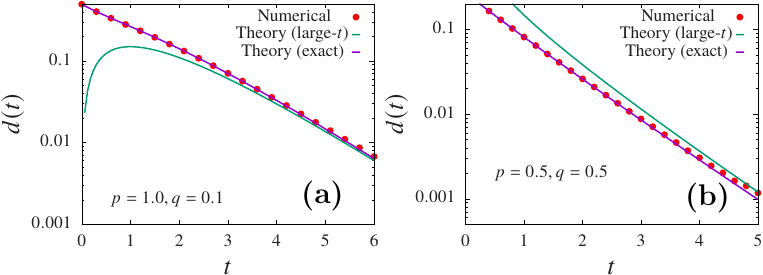}
	%\centering
	%  \subfigure{\includegraphics[scale=0.35]{avg-uniform.pdf}}
	% \subfigure{\includegraphics[scale=0.35]{avg-pos-power.pdf}}
	%\centering
	\caption{Comparison of the relaxation behaviour of the variance $\langle z_0^2(t) \rangle_c$ for the symmetric $(p=q)$ and asymmetric $(p \neq q)$ cases. For illustration, we have plotted $ d(t) = \langle z_0^2(t \to \infty) \rangle_c-\langle z_0^2(t) \rangle_c  $ in Eq. \eqref{eq-corr-eq-13} and compared it with the numerical simulation for two sets of parameters : $p=1.0~,q=0.1$ in panel (a) and $p=q=0.5$ in panel (b).
% The red circular points represent numerical result whereas the green solid lines corresponds to their approximate theoretical expressions. 
For both panels, we have performed simulation with $N=101$ particles, resetting rate $r=1$ and jump distribution $R (\eta)=1$.}    
	\label{msd-pic-2}
\end{figure*} 

%\begin{figure}[t]
%\includegraphics[scale=0.28]{asy-msd-uniform.pdf}
%\includegraphics[scale=0.29]{asy-msd-power.pdf}
%\centering
%\caption{Comparison of $ d_i(t) = \langle z_0^2(t \to \infty) \rangle_c-\langle z_0^2(t) \rangle_c  $ in Eq. \eqref{eq-corr-eq-13} with the numerical simulation for jump distribution $R(\eta) = 1$ [panel (a)] and $R(\eta) = 3(1-\eta)^2$ [panel (b)]. For both panels, we have chosen $p=1,~q=0.1$ and $r=1$.}    
%\label{msd-pic-2}
%\end{figure}
\subsection{Variance of $z_0(t)$}
\label{sec-var}
We first look at the variance for which we put $i=0$ in Eq. \eqref{eq-corr-eq-5}. Looking at this expression, it is clear that one needs to invert the Laplace transformation $\hat{\mathcal{G}}_0(s)$. We rewrite its expression from Eq. \eqref{eq-corr-eq-6} as
\begin{align}
\hat{\mathcal{G}}_0(s) =  \frac{\mu _1 \mu _2 (p+q) (\mu _1 - \mu _2)^{-1}(s \sqrt{s+r})^{-1}}{ \left[ \sqrt{(s+r) +4 \mu _1 (p+q)} +\frac{\mu _2}{\mu _1 - \mu _2}\sqrt{s+r}\right]}.
\label{eq-corr-eq-7}
\end{align}
Fortunately, one can perform this inversion exactly and we show in Appendix \ref{ILT-appen} that 
\begin{align}
\mathcal{G}_0(t) = &\frac{\mu _2(\mu _1 - \mu _2)(p+q)}{\sqrt{r}(\mu _1 - 2 \mu _2)}  \nonumber \\
& \times \int _{0}^{t}dT~ e^{- r T}~ \text{Erf}\left(\sqrt{r(t-T)} \right)~Z(T),
\label{eq-corr-eq-8}
\end{align}
where $\mathcal{G}_0(t)$ stands for the Laplace transformation of $\hat{\mathcal{G}}_0(s)$ and the function $Z(t)$ is defined as
\begin{align}
Z(t) =  &\ \frac{1}{\sqrt{\pi t}} \left[ e^{-4 \mu _1 t(p+q)}-\frac{\mu _2}{\mu _1 - \mu _2}\right] \nonumber  \\
& -\frac{\mu _2}{\mu _2 - \mu _2}~\sqrt{-B_1}~e^{-B_1 t}~ \text{Erf}(\sqrt{-B_1~ t}) \label{eq-corr-eq-9}  \\
& ~~~~~~~~~~~~~~~~~~~+ \sqrt{-B_2}~e^{-B_1 t}~ \text{Erf}(\sqrt{-B_2~ t}). \nonumber
\end{align}
Constants $B_1$ and $B_2$ depend on the model parameters as
 $B_1 = 4 (p+q)(\mu _1 - \mu _2)^2 /(\mu _1-2 \mu _2)$ and $B_2 = B_1- 4 \mu _1 (p+q)$. Finally, using Eq. \eqref{eq-corr-eq-8} in Eq. \eqref{eq-corr-eq-5}, we obtain
\begin{align}
C_0(t) = \mathcal{G}_0(t)+ \frac{2 \mu _1 ^2 (p-q)^2}{r^2} \left[1-e^{-rt}\left(rt+1 \right) \right],
\label{eq-corr-eq-10}
\end{align}
from which the variance of $z_0(t)$ turns out to be
\begin{align}
\langle z_0^2(t) \rangle _c & = C_0(t) -\langle z_0(t) \rangle^2, \nonumber \\
& = \mathcal{G}_0(t) +\frac{\mu _1 ^2 (p-q)^2}{r^2} \left[1-e^{-rt}\left(2rt+e^{-r t} \right) \right].
\label{eq-corr-eq-11}
\end{align}
This represents the exact variance of the position of a tagged particle in RAP with resetting. Figure \ref{msd-pic-1} illustrates the comparison of our analytical results with the numerical simulations for both symmetric and asymmetric cases. An excellent match is seen in both cases. For any non-zero $r$, we anticipate the expression in Eq.~\eqref{eq-corr-eq-11} to attain a stationary value at large times. To show this, one can, in principle, directly put $t \to \infty$ in Eq. \eqref{eq-corr-eq-8} and carry out the integration. However, it turns out more convenient to put $s \to 0$ limit in $\hat{\mathcal{G}}_0(s)$ and use the formula $\mathcal{G}_0 \left( t \to \infty  \right) = \lim _{s \to 0} \left[ s \hat{\mathcal{G}}_0(s) \right]$. With this procedure, the stationary value of the variance turns out to be
\begin{align}
 \langle z_0^2(t \to \infty) \rangle _c  =& \frac{\mu _1^2(p-q)^2}{r^2}+\frac{\mu _1 \mu _2 (p+q)}{\sqrt{r}(\mu _1 -  \mu _2)} \nonumber \\
& \times  \left[ \sqrt{r+4 \mu _1(p+q)}  + \frac{\mu _2 \sqrt{r}}{\mu _1 - \mu _2}\right]^{-1} . \label{eq-corr-eq-14}
\end{align}
Recall that for $r =0$, the tracer particle does not attain any stationary value. This indicates that the stationary value of the variance should diverge as $r \to 0$. From the exact expression in Eq.~\eqref{eq-corr-eq-14}, we find that $\langle z_0^2(t \to \infty) \rangle _c$ diverges differently depending on whether the particles move symmetrically $(p=q)$ or asymmetrically $(p \neq q)$. For $p=q$, the stationary value diverges as $\sim r^{-1/2}$ as $r \to 0$ whereas for $p \neq q$, it diverges as $\sim r^{-2}$. Contrarily for large $r$, $\langle z_0^2(t \to \infty) \rangle _c$ decays as $\sim 1/r$ for both cases.
 
\begin{figure*}[t]
	
	\includegraphics[width=\textwidth]{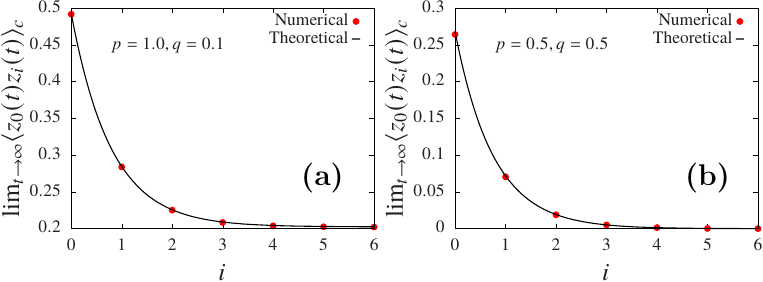}
	%\centering
	%  \subfigure{\includegraphics[scale=0.35]{avg-uniform.pdf}}
	% \subfigure{\includegraphics[scale=0.35]{avg-pos-power.pdf}}
	%\centering
	\caption{Comparison of the equal time correlation $ \langle z_0(t) z_i(t) \rangle_c  $ in the steady state $(t \to \infty)$ in Eq. \eqref{eq-corr-eq-20} for for two sets of parameters : $p=1.0,~q=0.1$ in panel (a) and $p=0.5,~q=0.5$ in panel (b). In both panels, simulation has been carried out with $N=101$ particles, resetting rate $r=1$ and jump distribution $R(\eta) = 1$.}   
%The red circular points represent numerical result whereas the black solid lines correspond to their theoretical expressions.  
	\label{corr-pic-0}
\end{figure*} 

After analysing the stationary value of the variance, let us look at its time-dependent form in Eq.~\eqref{eq-corr-eq-11}. 
%Let us now analyse its time-dependent expression in Eq. \eqref{eq-corr-eq-11}. 
While this is an exact expression, it has rather a complicated form. To gain some insights in it, we will analyse it for large $t$ and study the relaxation properties. For continuity of presentation, we have shown this calculation in Appendix \ref{ILT-appen} and quote only the final result here. We find  $d(t) =  \langle z_0^2(t \to \infty) \rangle _c -  \langle z_0^2(t) \rangle _c  $ is given by

\begin{align}
d(t) & \simeq \frac{\mu _2 \sqrt{\mu _1 p}}{ \sqrt{2\pi} r (\mu _1-\mu _2)} \frac{e^{-r t}}{\sqrt{t}},~~~~\text{for }p=q,  \label{eq-corr-eq-1345}\\
& \simeq \frac{2 \mu _1 ^2 (p-q)^2}{r^2} r t e^{-r t}, ~~~~~~~\text{for }p \neq q. \label{eq-corr-eq-13}
\end{align}
Interestingly, we find different relaxation behaviours depending on whether $p=q$ or $p \neq q$. While for the symmetric case, the variance relaxes as $\sim  e^{- r t} / \sqrt{t}$ to its stationary value, we obtain $\sim t e^{- r t} $ relaxation for the asymmetric case. Note that such difference between the symmetric and the asymmetric variances is not seen for free RAP $(r=0)$ and we obtain the same sub-diffusive scaling for both cases as shown in Eq.~\eqref{msd-no}. Indeed, writing the time evolution equation for the variance $\langle z_0^2(t) \rangle _c$ for the free RAP, one can show that it can be made independent of $p$ and $q$ by suitably scaling $t \to (p+q)t$ \cite{PhysRevE.64.036103}. Therefore, we get same temporal scaling of the variance for both symmetric and the asymmetric cases. However, in presence of resetting, we get an additional time scale in the problem $\left( \sim 1/r \right)$ and the time evolution equation cannot be rendered independent of $p$ and $q$. This gives rise to different behaviours for two cases. This key difference is one of the consequences of the resetting. {In Figure \ref{msd-pic-2}, we have compared the relaxation properties with the numerical simulations for $p=q$ case (right panel) and $p \neq q$ case (left panel). We observe an excellent agreement between our theory and numerics for both cases.}
%both the exact expression and the asymptotic expressions of $\langle z_0^2(t) \rangle _c$ with the numerical simulation. We observe excellent agreement  between theory and numerics for all cases.

%\begin{figure}[t]
%\includegraphics[scale=0.28]{eq-corr-uniform.pdf}
%\includegraphics[scale=0.3]{eq-corr-st-uniform.pdf}
%\centering
%\caption{(a) Comparison of $ \langle z_0(t) z_i(t) \rangle_c  $ in Eq. \eqref{eq-corr-eq-19} with the numerical simulation for different values of $t$. For each $t$, the solid line represents the analytical formula in Eq. \eqref{eq-corr-eq-19} while circles represent simulation data. For simplicity, we have shown the comparison only for $i \geq 1$. We have chosen $p=0.5,~q=0.2$ and $r=0.2$. (b) Comparison of the correlation $ \langle z_0(t) z_i(t) \rangle_c  $ in the steady state in Eq. \eqref{eq-corr-eq-20} for $p=0.5,~q=0.2$ and $r=1$. For both panels, the jump distribution $R(\eta)$ is taken to be uniform.\textcolor{red}{\textbf{Due}}}    
%\label{corr-pic-1}
%\end{figure}
\subsection{Correlation $C_i(t) = \langle z_0(t) z_i(t)\rangle $ for $i \neq 0$}
\label{sec-corr}
After analysing the variance of the position of a tagged particle, let us now look at the position correlation for two different tagged particles. For free RAP, we saw in Eq. \eqref{corr-no} that the two-point correlation function satisfies non-trivial scaling behaviour in $\sim |i| / \sqrt{t}$ with $i$ being the separation between two particles. The associated scaling function is given in Eq. \eqref{scaling-no}. In this section, we study the correlation function in presence of resetting and calculate $C_i(t) = \langle z_0(t) z_i(t) \rangle $ for $i \neq 0$. Its expression in the Laplace domain is given in Eq. \eqref{eq-corr-eq-5} from which it is clear that we have to perform the inverse Laplace transformation of $\hat{\mathcal{G}}_i(s)$. Rewriting its expression from Eq. \eqref{eq-corr-eq-6}
\begin{align}
& \hat{\mathcal{G}}_i(s) = \hat{\mathcal{G}}_0(s) \times \hat{w}_{i}(s+r), \label{eq-corr-eq-15} \\
\text{with     }& \hat{w}_{i}(s) =  \left[ \frac{s +2 \mu _1' -\sqrt{s^2 + 4 \mu _1 ' s} }{2 \mu _1'}\right]^{|i|}, 
\label{new-pssfaf-eq-1}
\end{align}
where we have defined $\mu _1' = \mu _1(p+q)$.  In order to perform the inverse Laplace transformation of  $\hat{\mathcal{G}}_i(s)$, we have to carry out two inversions: one is for $\hat{\mathcal{G}}_0(s)$ and the other is for $\hat{w}_{i}(s)$. Now for  $\hat{\mathcal{G}}_0(s)$ , we have already computed this inversion in Eq.~\eqref{eq-corr-eq-8}. On the other hand, for $\hat{w}_{i}(s)$, its inversion in the time domain (denoted by $w_i(t)$) is given by \cite{bateman_2023_mhd23-e0z22}
\begin{align}
w_i(t) = \frac{|i|}{t}~e^{-2\mu _1't}~I_{|i|}\left( 2 \mu _1 ' t \right),~~~\text{for } i \neq 0.
\label{eq-corr-eq-17}
\end{align}
Finally using the convolution form of $\hat{\mathcal{G}}_i(s)$ in Eq. \eqref{eq-corr-eq-15} gives
\begin{align}
\mathcal{G}_i(t) = \int _0^{t} d T ~e^{-r T}~ w_i(T)~ \mathcal{G}_0(t-T).
\label{eq-corr-eq-16}
\end{align}
inserting which in Eq. \eqref{eq-corr-eq-5}, we obtain
\begin{align}
C_i(t) = &  
 \int _0^{t} d T ~e^{-r t}~ w_i(T)~ \mathcal{G}_0(t-T) \nonumber \\
& ~~~~~~+ \frac{2 \mu _1 ^2 (p-q)^2}{r^2} \left[1-e^{-rt}\left(rt+1 \right) \right].
\label{eq-corr-eq-18}
\end{align}
Remember that $C_i(t) = \langle z_0(t) z_i(t) \rangle$ and in order to obtain the connected-correlation, we subtract the mean contributions as follows
\begin{align}
\langle z_0(t) z_i(t) \rangle _c & = \langle z_0(t) z_i(t) \rangle-\langle z_0(t) \rangle \langle z_i(t) \rangle, \nonumber \\
& = \int _0^{t} d T ~e^{-r t}~ w_i(T)~ \mathcal{G}_0(t-T) \label{eq-corr-eq-19} \\
&~~~~~  +\frac{\mu _1 ^2 (p-q)^2}{r^2} \left[1-e^{-rt}\left(2rt+e^{-r t} \right) \right]. \nonumber
\end{align}
It is worth mentioning that even for free RAP, only asymptotic results for $\langle z_0(t) z_i(t) \rangle _c$ are known \bluew{\cite{PhysRevE.64.036103}}. Our analysis here provides an exact expression for the correlation valid for all times and not just at large times. However, at large times, one can simplify this expression  In fact, putting $t \to \infty$ in Eq. \eqref{eq-corr-eq-16}, one can see that
\begin{align}
\mathcal{G}_i(t \to \infty) = \mathcal{G}_0 \left( t \to \infty \right) ~\hat{w}_i(r),
\end{align}
where the steady value $\mathcal{G}_0 \left( t \to \infty \right)$ is given in Eq.~\eqref{psdf-eq} and $\hat{w}_i(r)$ is given in Eq.~\eqref{new-pssfaf-eq-1}. Using these expressions, we find that the correlation between two tracer particles in the steady state is given by
\begin{align}
\langle z_0(t) z_i(t) \rangle _c ~\stackrel{t \to \infty}{=}~ \mathcal{G}_0 ^{\text{st}} ~\text{Exp}\left(-\frac{|i|}{\xi_p}\right)+\frac{\mu _1^2(p-q)^2}{r^2},
\label{eq-corr-eq-20}
\end{align}
where $\mathcal{G}_0 ^{\text{st}} = \mathcal{G}_0 (t \to \infty)$ 
is defined in Eq. \eqref{psdf-eq} and the decay length $\xi_p$ is defined as
\begin{align}
\xi_p =  |\log \left(r+2 \mu _1'-\sqrt{r^2+4 \mu _1' r} \right)-\log(2 \mu _1')|^{-1} .
\label{eq-corr-eq-21}
\end{align}
Interestingly for $p \neq q$, we find that the correlation decays to a non-zero constant value as $\big| i \big| \to \infty$. Contrarily, it decays to zero for $p=q$. For reset-free RAP, this correlation decays to zero both for the symmetric and the asymmetric cases. Our study unravels that resetting affects these two cases in different manners and manifestly gives rise to long-range correlations only for the $p \neq q$ case. Recently, resetting induced long-range correlation was also found in independently diffusing particles but subjected to simultaneous resetting at a rate $r$ \cite{PhysRevLett.130.207101}. Here, we have extended this result for interacting single-file systems. Physically, the long-range correlation can be understood as follows: Consider two particles with positions $x_0(t)$ and $x_l(t)$ with $l \gg 1$. Both these particles experience an effective attraction around their initial positions due to the resetting event. However, this attraction has a purely dynamical interpretation and does not arise due to any physical potential. Furthermore, for $p > q$, all particles that lie between $x_0(t)$ and $x_l(t)$ experience a net drift towards $x_l(t)$. As a result of this combined effect of attraction and drift, the positions $x_0(t)$ and $x_l(t)$ of two particles become strongly correlated. Figure \ref{corr-pic-0} shows the comparison of our analytical results with the same obtained using numerical simulation. Indeed, even in simulations, we find that the correlation $C_{i}(t)$ does not decay to zero for the asymmetric case.

\section{Unequal time correlations}
So far, we have presented rigorous results on the variance and the equal time correlation and demonstrated how resetting modifies these quantities. Our analysis showed that contrary to the reset-free RAP model, these quantities, in presence of resetting, behave differently depending on the presence or absence of drive in the dynamics. Continuing on this, we now look at the autocorrelation and the unequal time position correlations for two tracer particles. Let us denote this correlation by $S_i(t_0,t_0+t)=\langle z_0(t_0) z_i(t_0+t) \rangle$. As done before, we again consider a small time interval $[t_0+t, t_0+t + dt]$ and follow the update rules in Eqs.~\eqref{update-rule-3} and \eqref{update-rule-4} to write down the total change in $S_i(t_0,t_0+t)$ within this interval. This results in the following differential equation:
\begin{figure*}[t]	
	\includegraphics[width=\textwidth]{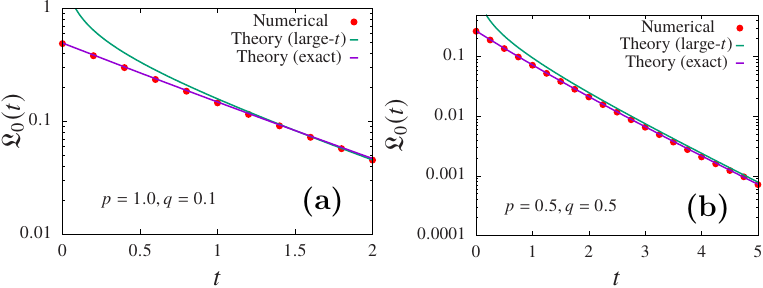}
	%\centering
	%  \subfigure{\includegraphics[scale=0.35]{avg-uniform.pdf}}
	% \subfigure{\includegraphics[scale=0.35]{avg-pos-power.pdf}}
	%\centering
	\caption{Autocorrelation $\mathcal{L}_0(t)$ has been compared with the numerical simulations for $p=1.0,q=0.1$ in panel (a) and $p=0.5,q=0.5$ in panel (b). We have plotted both the exact expression in Eq.~\eqref{uneq-psss-eq-3} as well as the asymptotic (large-$t$) expression in Eq.~\eqref{uneq-psss-eq-5}. Simulation has been performed with $N=101$ particles, resetting rate $r=1$ and jump distribution $R(\eta)=1$.}   
	\label{auto_corr-pic-0}
\end{figure*} 
\begin{align}
\frac{dS_i(t_0,t_0+t)}{dt}= &-r S_i +\mu_1 (p-q) h(t_0) \label{eq:uneq_time_evln} \\
&+\mu_1 \Big[ p S_{i+1}+q S_{i-1}  -(p+q) S_i\Big]. \notag
\end{align}
Solving this equation by taking the joint Fourier-Laplace transformations as
%\normalsize{The} Eq.~\eqref{eq:uneq_time_evln} is linear in $S_i$, thus we proceed by taking  Fourier-transformation
\begin{align}
& \tilde{S}(k,t_0,t_0+t) =\sum_{i=-\infty}^{\infty} e^{jik} ~S_i(t_0,t_0+t), \\
& \mathbb{S}(k,s,t) = \int _{0}^{\infty}dt_0 ~e^{-s t_0}~ \tilde{S}(k,t_0,t_0+t), \label{ghdcod-eq-2}
\end{align}
and inserting them in Eq.~\eqref{eq:uneq_time_evln}, we obtain
\begin{align}
\frac{d \mathbb{S}}{dt}=-\alpha(k) \mathbb{S}+2 \pi \mu_1 (p-q) \hat{h}(s) \delta(k), \label{eq:uneq_time_evln_FT}
\end{align}
%\footnotesize{\begin{align}
%\frac{d \tilde{S}(k,t_0,t_0+t)}{dt}=2 \pi \mu_1 (p-q) h(t_0) \delta(k) -\alpha(k) \tilde{S}(k,t_0,t_0+t),
%\label{eq:uneq_time_evln_FT}
%\end{align}}
where $\alpha(k)=r+\mu _1(p+q)-\mu_1 p e^{j k}-\mu_1 q e^{-j k}$. Since, we are interested in computing the correlations in the steady state $(\text{i.e. }t_0 \to \infty)$, we will analyse Eq.~\eqref{eq:uneq_time_evln_FT} in the small-$s$ limit. In Appendix \ref{appen-unequal}, we have explicitly carried out this analysis and obtained the correlation $S_i \left(t_0, t_0+t \right)$ measured from the steady-state $(t_0 \to \infty)$ as
%Solving this equation and performing the inverse Fourier transformation for $t_0 \to \infty$ [see Appendix \ref{appen-unequal}], we obtain
\begin{align}
S_i \left(t_0, t_0+t \right)& \simeq~ \frac{\mathcal{B} r}{2 \pi} \int _{-\pi} ^{\pi}dk ~\frac{e^{-j i k - \alpha(k)t}}{r+2 \mu _1' (1-\cos k)} \nonumber \\
& ~~~+ \frac{\mu _1 ^2(p-q)^2}{r^2}(1+e^{-r t}) \label{uneq-psss-eq-1}\\
\text{with }~~\mathcal{B} & =\sqrt{ \frac{r+ 4 \mu _1'}{r}}~\mathcal{G}_0 ^{\text{st}}, \label{uneq-psss-eq-2}
\end{align}
where again we have used the notation $\mu _1 ' = \mu_1(p+q) $ and $\mathcal{G}_0 ^{\text{st}} = \mathcal{G}_0 (t_0 \to \infty)$ is defined in Eq. \eqref{psdf-eq}. Subtracting the mean contribution from this correlation, we obtain the connected-correlation $\mathcal{L}_i(t)$ as
\begin{align}
\mathcal{L}_i(t) & = \Big[S_i \left(t_0, t_0+t \right)-\langle z_0(t_0) \rangle ~\langle z_i(t_0+t) \rangle \Big] _{t_0 \to \infty} , \nonumber \\
& = \frac{\mathcal{B} r}{2 \pi} \int _{-\pi} ^{\pi}dk ~\frac{e^{-j i k - \alpha(k)t}}{r+2 \mu _1' (1-\cos k)}+ \frac{\mu _1 ^2(p-q)^2}{r^2} e^{-r t}. \label{uneq-psss-eq-3}
\end{align}
For $t=0$, one can perform the integration over $k$ and the result matches with the equal time correlation in Eq.~\eqref{eq-corr-eq-20} in the steady state. On the other hand, for non-zero $t$, performing this integration turns out to be difficult. However, for large $t$, we could carry out the integration rigorously and obtain some asymptotic results. Below we discuss this first for the autocorrelation $(i=0)$ and then for general $i$.
\begin{figure*}[t]	
	%\includegraphics[width=\textwidth]{figure6.pdf}
	%\centering
	  \subfigure{\includegraphics[scale=0.3]{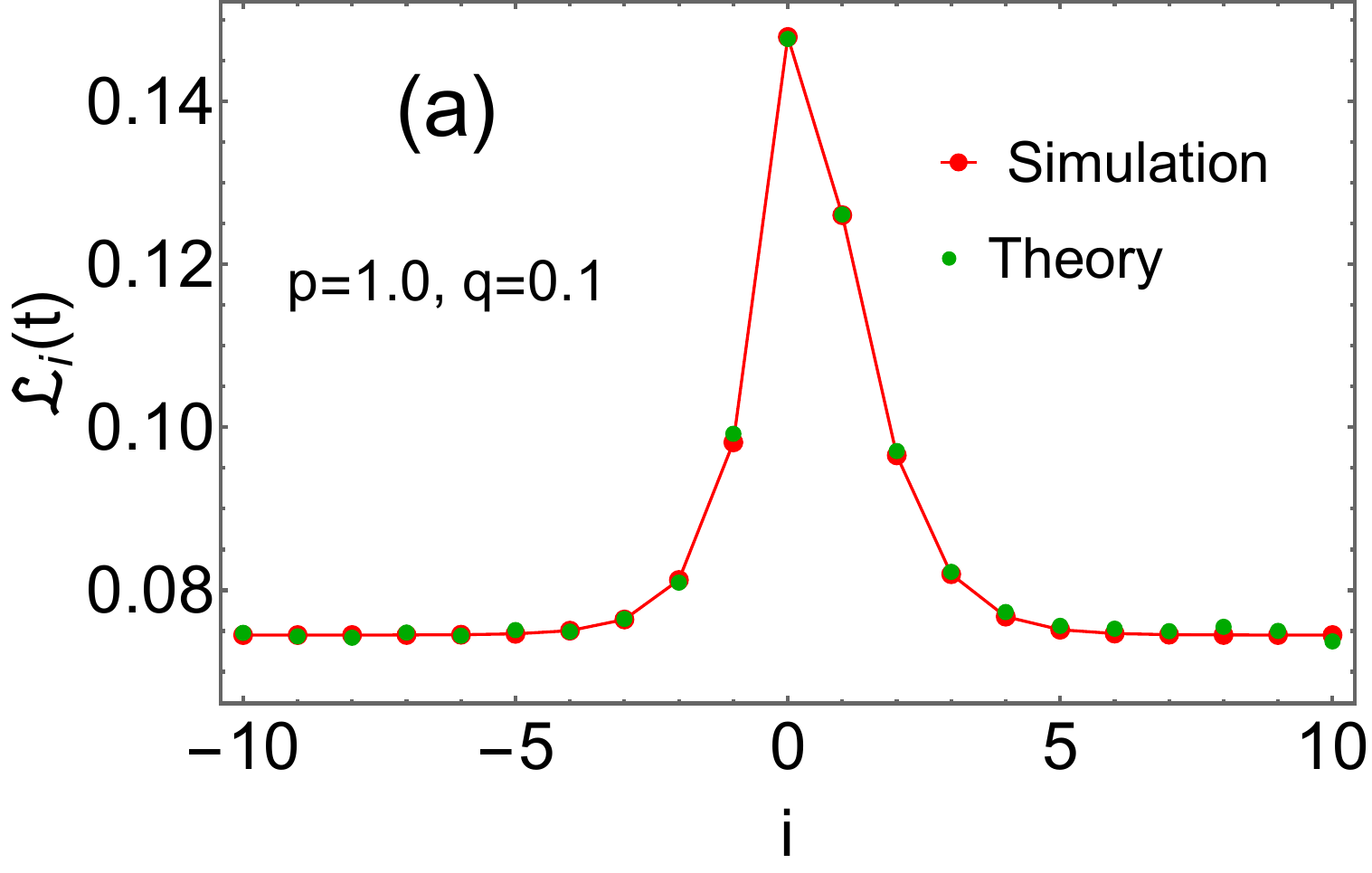}}
	 \subfigure{\includegraphics[scale=0.3]{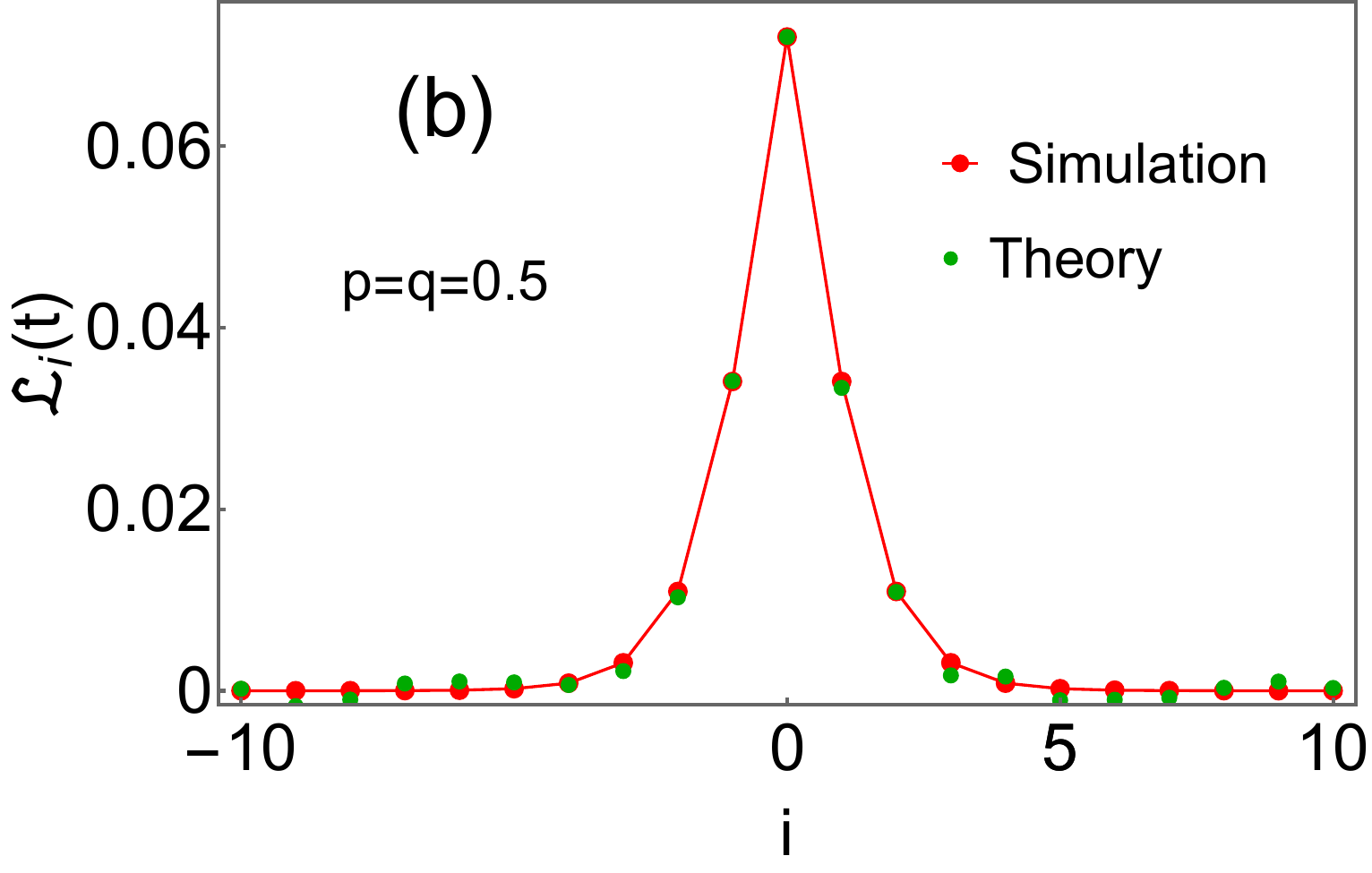}}
	%\centering
	\caption{Unequal time correlation $\mathcal{L}_i(t)$ in Eq.~\eqref{uneq-psss-eq-3} is compared with the numerical simulation for asymmetric ($p>q$) RAP in the left panel and symmetric ($p=q$) RAP in the right panel for $t=1$. Parameters chosen are $r=1$, $R(\eta)=1$ and number of particles $N=101$ for simulation.}    
	\label{uneq-fig-new-1}
\end{figure*}

\subsection{Autocorrelation $\mathcal{L}_0(t)$}
\label{sec-auto}
To get large $t$-behaviour of $\mathcal{L}_0(t)$, we first notice that one gets exponentially decaying terms like $\sim \exp\left[-2 \mu _1(p+q)t \sin ^2(k/2)\right]$ inside the integration in Eq.~\eqref{uneq-psss-eq-3}. For large $t$, such integrations will be dominated by smaller values of $k$. Therefore, performing small-$k$ approximation in Eq.~\eqref{uneq-psss-eq-3}, we get
\begin{align}
\mathcal{L}_0(t) \simeq & \frac{\mathcal{B}e^{-r t}}{2 \pi}\int _{-\pi}^{\pi} dk~e^{j k \mu _1(p-q)t-\mu _1 t (p+q)k^2/2} \nonumber \\
&~~~~~~~~~~~~~~~~~+ \frac{\mu _1 ^2(p-q)^2}{r^2} e^{-r t}. \label{uneq-psss-eq-4}
\end{align}
Next, we change the variable $u = k \sqrt{\mu _1 (p+q)t}$ in this equation and carry out the integration for large $t $ to obtain
\begin{align}
\mathcal{L}_0(t) \simeq \frac{\mathcal{B} ~e^{- \left[ r+\frac{\mu_1(p-q)^2}{2(p+q)}\right]t}}{\sqrt{2 \pi \mu _1(p+q)t}}+\frac{\mu _1 ^2(p-q)^2}{r^2} e^{-r t}. \label{uneq-psss-eq-5}
\end{align}
Using this expression, we again find that the large time decay of $\mathcal{L}_0(t)$ depends on whether the particles experience a drift or not. While for the symmetric case, the autocorrelation in the steady state decays as $\sim e^{- r t}/ \sqrt{t}$ at late times, we observe an exponential decay $\sim e^{-r t}$ for the asymmetric $(p \neq q)$ case. This is contrary to the case of simple resetting Brownian motion, where autocorrelation in the steady state decays exponentially as $e^{-rt}$ at all times \cite{Stojkoski_2022_auto, Majumdar_2018_auto}. However, for interacting particles, $\mathcal{L}_0(t)$ has a more complicated form and picks up exponential decay (or otherwise) only at large times. In Figure \ref{auto_corr-pic-0}, we have compared this late time decay with the numerical simulations. We observe that while the simulation data deviates from Eq.~\eqref{uneq-psss-eq-5} at small times, the agreement becomes better at larger times.

\subsection{Unequal time correlation $\mathcal{L}_i(t)$}
\label{sec-uneq}
We now analyse Eq.~\eqref{uneq-psss-eq-3} for general $i$. For this case also, we can perform small-$k$ approximation in Eq.~\eqref{uneq-psss-eq-3} for larger values of $t$ since the integral has exponentially decaying terms like  $\sim \exp\left[-2 \mu _1(p+q)t \sin ^2(k/2)\right]$. We then obtain
\begin{align}
\mathcal{L}_i(t) \simeq & \frac{\mathcal{B}e^{-r t}}{2 \pi}\int _{-\pi}^{\pi} dk~e^{-jik+j k \mu _1(p-q)t-\mu _1 t (p+q)k^2/2} \nonumber \\
&~~~~~~~~~~~~~~~~~+ \frac{\mu _1 ^2(p-q)^2}{r^2} e^{-r t}. \label{uneq-psss-eq-6}
\end{align}
To perform the integration over $k$, we change the variable $u = k \sqrt{\mu _1 (p+q)t}$ and plug it in this equation. Finally, we find that the $\mathcal{L}_i(t)$ satisfies the following scaling relation
\footnotesize{\begin{align}
\mathcal{L}_i(t) - \mathcal{L}_{|i| \to \infty}(t) \simeq \frac{\mathcal{B} ~e^{- rt}}{\sqrt{2 \pi \mu _1(p+q)t}}~ \mathcal{M}\left( \frac{i-\mu _1(p-q)t}{\sqrt{2 \mu _1(p+q)t}} \right),\label{uneq-psss-eq-7}
\end{align}}
\normalsize{where} the scaling function $\mathcal{M}(y)$ and $\mathcal{L}_{|i| \to \infty}(t)$ are given by
\begin{align}
\mathcal{M}(y) = e^{-y^2}, ~\text{and } ~~\mathcal{L}_{|i| \to \infty}(t) = \frac{\mu _1 ^2(p-q)^2}{r^2} e^{-r t}. \label{uneq-psss-eq-8}
\end{align}
Note that this scaling behaviour is entirely an outcome of the resetting dynamics and does not appear for the reset-free RAP \cite{PhysRevE.64.036103}. Looking at Eq.~\eqref{uneq-psss-eq-7}, once again we see that for the asymmetric RAP, $\mathcal{L}_i(t)$ takes a non-zero value as $|i| \to \infty$ indicating a long-range correlation between two particles. However, this value decays exponentially with time and as $t \to \infty$, this long-range correlation vanishes. As discussed in the case of equal time correlation, the appearance of long-range correlation turns out to be an interplay of the effective attraction experienced by the particles around their resetting sites and a net drift due to asymmetric rates.

\begin{figure}[t]
\includegraphics[scale=0.26]{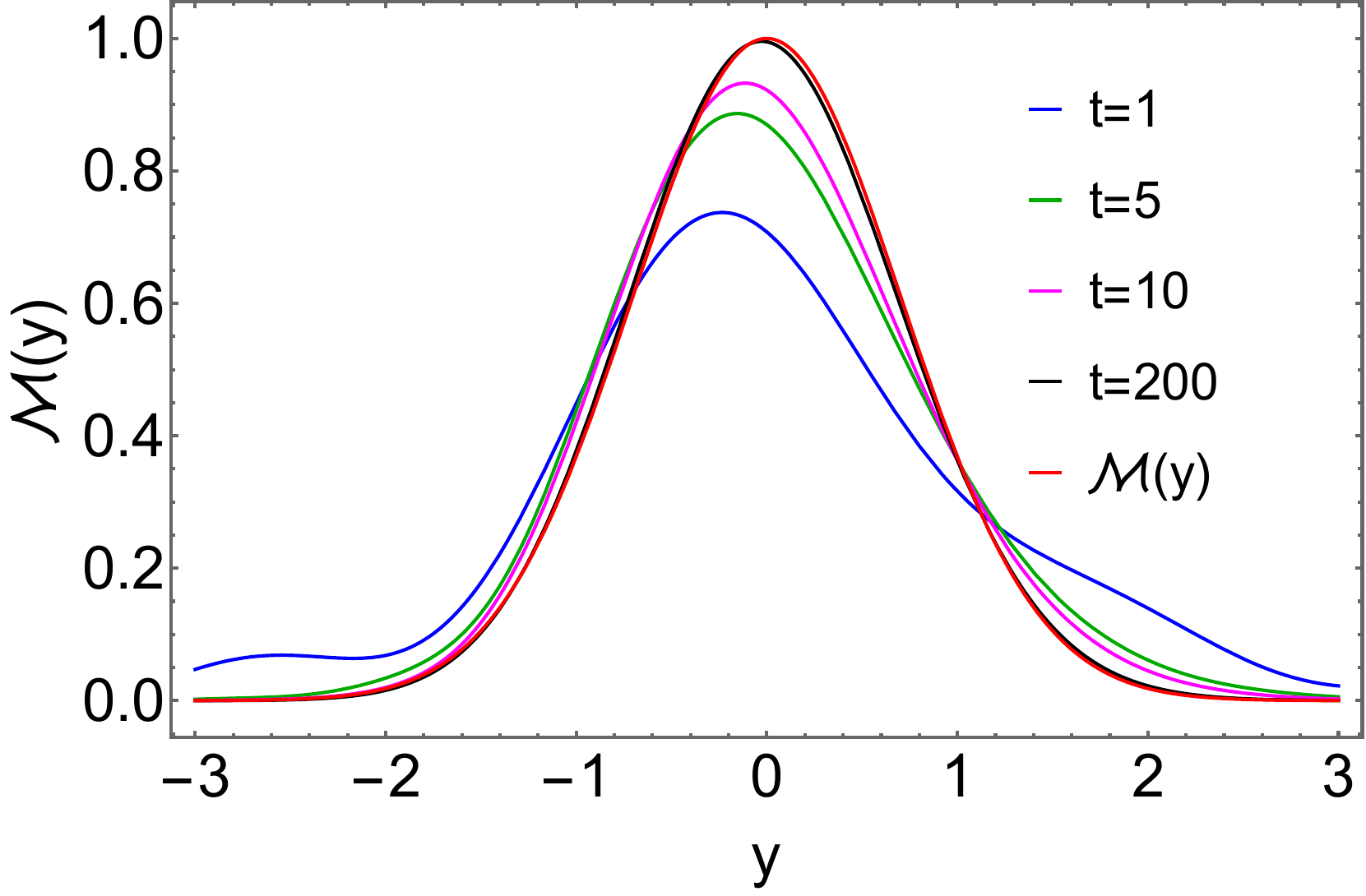}
\centering
\caption{Scaling function $\mathcal{M}(y)$ in Eq.~\eqref{uneq-psss-eq-7} for the unequal time correlation $\mathcal{L}_i(t)$ is compared with the exact expression of $\mathcal{L}_i(t)$ in Eq.~\eqref{uneq-psss-eq-3} for different values of $t$ and  $p=1.0,~q=0.1,~r=1$ and jump distribution $R(\eta)=1$.}
\label{uneq-fig-new-2}
\end{figure}

In Figure \ref{uneq-fig-new-1}, we have compared the exact expression of $\mathcal{L}_i(t)$ in Eq.~\eqref{uneq-psss-eq-3} with the numerical simulations for $p \neq q$ in the left panel and $p=q$ in the right panel. For both cases, we see an excellent agreement between theory and numerics. However, demonstrating the scaling behaviour of $\mathcal{L}_i(t)$ in Eq.~\eqref{uneq-psss-eq-7} in simulation turns out to be difficult. It turns out that one needs to go to very large values of $t$ in order to observe this scaling relation. For instance, in Figure \ref{uneq-fig-new-2}, we see that this scaling behaviour becomes valid at around $t=200$. However, the value of $\mathcal{L}_i(t)$ at such large times is very small due to the presence $\sim e^{-r t}$ term in Eq.~\eqref{uneq-psss-eq-7}. Measuring such small values in simulation is difficult. Therefore, to validate this scaling relation, we have plotted the exact $\mathcal{L}_i(t)$ in Figure \ref{uneq-fig-new-2} for different values of $t$ by numerically performing the integration over $k$. At large $t$, we recover the scaling function $\mathcal{M}(y)$ [see Figure \ref{uneq-fig-new-2}].

\section{Conclusion}
\label{sec-conclusion}
In conclusion, we have studied the motion of tracer particles in a one dimensional single-file model called random average process which is subjected to stochastic resetting. The resetting mechanism, characterized by a constant rate $r$ causes the entire system being reinstated to the configuration given in Eq.~\eqref{ini-pos}. Utilizing an exact microscopic analysis, we calculated key statistical quantities such as variance, equal-time correlation, autocorrelation, and unequal time correlation for the positions of tracer particles. Through these calculations, we demonstrated how resetting modifies the system and gives rise to properties which are otherwise not observed in absence of the resetting

We first looked at the variance $\langle z_0^2(t) \rangle _c$ whose exact expression is given in Eq.~\eqref{eq-corr-eq-11}. At large times, it expectedly attains a stationary value given in Eq.~\eqref{eq-corr-eq-14}. To gain some physical insights, we further explored the relaxation behaviour of $\langle z_0^2(t) \rangle _c$ as it approaches its stationary value. Interestingly, this relaxation process turns out to crucially depend on whether the particles move symmetrically $(p=q)$ or asymmetrically $(p \neq q)$ on either side. While for the symmetric case, the variance relaxes as $\sim  e^{- r t} / \sqrt{t}$ to its stationary value, we obtain $\sim t e^{- r t} $ relaxation for the asymmetric case. Note that such difference between the symmetric and the asymmetric variances is not seen for free RAP $(r=0)$ and we obtain the same sub-diffusive scaling for both cases as shown in Eq.~\eqref{msd-no}. Resetting introduces an additional time scale in the model which leads to different behaviours for two cases. This key difference is one of the consequences of the resetting.

We next turned our attention to the equal-time position correlation for two different tracer particles. Focussing on the steady state, our study revealed that resetting induces a long-range correlation only for the asymmetric case. On the other hand, for the symmetric case, we obtained correlations that decay exponentially with the distance in Eq.~\eqref{eq-corr-eq-20}. In a recent work involving independently diffusing particles, but undergoing global resetting at a rate $r$, the authors showed that the simultaneous resetting induces a strong long-range correlation in the system \bluew{\cite{PhysRevLett.130.207101}}. Our work generalises these results in the interacting single-file set-up and shows that simultaneous resetting induces a long-range correlation only when particles experience a bias $(p \neq q)$.

%Finally, we investigated  the unequal time position correlation in the steady state, 
%$\left(i.e.~\footnotesize{\mathcal{L}_i(t) = \langle z_0(t_0) z_{i}(t_0+t) \rangle - \langle z_0(t_0)\rangle \langle z_i(t_0+t)\rangle \Big|_{t_0 \to \infty}}\right)$

Finally, we investigated the autocorrelation and the unequal time position correlation in the steady state. For the autocorrelation $\mathcal{L}_0(t)$, once again, we find that the large-$t$ decay is different for $p=q$ and $p \neq q$ cases. Specifically, for the symmetric case $\mathcal{L}_0(t)$ exhibited a decay of  $ \sim e^{- r t}/ \sqrt{t}$, while for the asymmetric case, the decay followed $\mathcal{L}_0(t) \sim e^{- r t}$. Conversely, the unequal time position correlation $\mathcal{L}_i(t)$ exhibits a scaling behaviour in terms of the variable $y=\big( i - \mu_1 (p-q) t\big)/\sqrt{t}$. The associated scaling function $\mathcal{M}(y)$ is written in Eq.~\eqref{uneq-psss-eq-8}. We emphasize that this scaling behaviour is entirely an outcome of the resetting dynamics and does not appear for the reset-free RAP \cite{PhysRevE.64.036103}.

Studying analytically an interacting multi-particle system is difficult because of the correlation between different particles. Here, we presented a specific single-file model for which exact microscopic computations can be carried out.
Our work pointed at a crucial difference in the tracer dynamics for symmetric RAP and asymmetric RAP, both subjected to resetting at a rate $r$. For future direction, it would be interesting to explore an intermediate case where only some of the particles experience bias while all others move symmetrically \cite{Cividini_2016_RAP, Kundu_2016_RAP} and see if one still gets a resetting induced long-range correlation. Also, our paper focused on one specific model of single-file motion called random average process. It remains a promising direction to study effects of resetting on other single-file models like single-file diffusion \bluew{\cite{Krapivsky_2015, SadhuPRL2014, Benichou_2018, PhysRevB.18.2011}}, in active particles \cite{Singh_2021_cross, Put_2019, Santra2023RAP, Teomy_2019, Galanti2013, D0SM00687D, Banerjee_2022} and also in experiments \bluew{\cite{reset-Exp1, reset-Exp2, reset-Exp3, reset-Exp3}}. Finally, we have only looked at different two-point correlation functions in our paper. Obtaining higher moments and the distribution function for the position of a tracer particle still remains an open problem even for the reset-free RAP.

\begin{acknowledgements}
We thank Arnab Pal and R. K. Singh for their useful comments on the paper. SS acknowledges the support of the Department of Atomic Energy, Government of
India, under Project No. 19P1112$\&$D. PS acknowledges the support of Novo Nordisk Foundation under the grant number NNF21OC0071284.
\end{acknowledgements}

\appendix
\begin{widetext}
\section{Derivation of the variance $\langle z_0^2(t) \rangle _c $ in Eq. \eqref{eq-corr-eq-11}}
\label{ILT-appen}
In this appendix, we provide a detailed derivation of the variance $\langle z_0^2(t) \rangle _c = \langle z_0^2(t) \rangle - \langle z_0(t) \rangle ^2$ in Eq. \eqref{eq-corr-eq-11}. The starting point is to find $C_0(t) = \langle z_0^2(t) \rangle$ for which we need the Laplace transform $\hat{C}_0(s)$ in Eq. \eqref{eq-corr-eq-5}. Rewriting this expression here as
\begin{align}
& \hat{C}_0(s)  = \frac{2 \mu _1^2 (p-q)^2}{s(s+r)^2}  + \hat{\mathcal{G}}_0(s), \label{appen-eqqq-1}
\end{align}
with $\hat{\mathcal{G}}_0(s) $ defined as
%\begin{align}
%& \hat{\mathcal{G}}_0(s) =  \frac{\mu _2(\mu _1 - \mu _2)(p+q)}{(\mu _1 - 2 \mu _2)}  \hat{H}(s) \hat{Z}(s+r),
%\label{eq-ILT-appen-1} \\
%& \hat{Z}(s) = \frac{\sqrt{s+4 \mu _1 (p+q)}-\frac{\mu _2}{\mu _1 - \mu _2} \sqrt{s}}{s+ \frac{4 (p+q)(\mu _1-\mu _2)^2}{\mu _1-2 \mu_2}}, \label{eq-ILT-appen-2} \\
%& \hat{H}(s) = \frac{1}{s \sqrt{s+r}}. \label{eq-ILT-appen-3} 
%\end{align}
\begin{align}
& \hat{\mathcal{G}}_0(s) =  \frac{\mu _2(\mu _1 - \mu _2)(p+q)}{(\mu _1 - 2 \mu _2)}  \hat{H}(s) \hat{Z}(s+r),~~~\text{with }
\hat{Z}(s) = \frac{\sqrt{s+4 \mu _1 (p+q)}-\frac{\mu _2}{\mu _1 - \mu _2} \sqrt{s}}{s+ \frac{4 (p+q)(\mu _1-\mu _2)^2}{\mu _1-2 \mu_2}}, \text{ and }\hat{H}(s) = \frac{1}{s \sqrt{s+r}}. \label{eq-ILT-appen-3-ps} 
\end{align}
To evaluate first term in the right hand side of Eq. \eqref{appen-eqqq-1}, we use the relation 
\begin{align}
\int _{0}^{\infty} dt ~e^{-s t} \left[\frac{1-e^{-r t} (rt+1)}{r^2} \right] = \frac{1}{s(s+r)^2}.
\end{align}
On the other hand, for second term, we use the convolution structure of $\hat{\mathcal{G}}_0(s) $ which gives $\mathcal{G}_0(t)$ as
\begin{align}
\mathcal{G}_0(t) = &\frac{\mu _2(\mu _1 - \mu _2)(p+q)}{(\mu _1 - 2 \mu _2)}   \int _0 ^{t} dT~e^{-r T} ~H(t-T)~ Z(T).
\label{eq-ILT-appen-3}
\end{align}
Here $H(t)$ and $Z(t)$ are inverse Laplace transforms of $\hat{H}(s)$ and $\hat{Z}(s)$ respectively. Using Mathematica, we find $H(t) =  \text{Erf} \left( \sqrt{rt} \right)/\sqrt{r}$ and $Z(t)$ as
\begin{align}
Z(t) =  &\ \frac{1}{\sqrt{\pi t}} \left[ e^{-4 \mu _1 t(p+q)}-\frac{\mu _2}{\mu _1 - \mu _2}\right] -\frac{\mu _2}{\mu _2 - \mu _2}~\sqrt{-B_1}~e^{-B_1 t}~ \text{Erf}(\sqrt{-B_1~ t})   + \sqrt{-B_2}~e^{-B_1 t}~ \text{Erf}(\sqrt{-B_2~ t}),\label{eq-ILT-appen-4}
\end{align}
with $B_1 = 4 (p+q)(\mu _1 - \mu _2)^2 /(\mu _1-2 \mu _2)$ and $B_2 = B_1- 4 \mu _1 (p+q)$. Substituting these expressions in Eq. \eqref{eq-ILT-appen-3} yields
\begin{align}
\mathcal{G}_0(t) =& \frac{\mu _2(\mu _1 - \mu _2)(p+q)}{\sqrt{r}(\mu _1 - 2 \mu _2)}   \int _{0}^{t}dT~ e^{- r T}~ \text{Erf}\left(\sqrt{r(t-T)} \right)~Z(T).
\label{eq-ILT-appen-5}
\end{align}
We now have both terms from the right hand side of Eq. \eqref{appen-eqqq-1}. Performing the inverse Laplace transformation, we get
\begin{align}
C_0(t) = \mathcal{G}_0(t)+ \frac{2 \mu _1 ^2 (p-q)^2}{r^2} \left[1-e^{-rt}\left(rt+1 \right) \right],
\end{align}
from which the variance of $z_0(t)$ turns out to be
\begin{align}
\langle z_0^2(t) \rangle _c & = C_0(t) -\langle z_0(t) \rangle^2, \nonumber \\
& = \mathcal{G}_0(t) +\frac{\mu _1 ^2 (p-q)^2}{r^2} \left[1-e^{-rt}\left(2rt+e^{-r t} \right) \right],
\label{eq-12corr-eq-11}
\end{align}
with $\mathcal{G}_0(t)$ given in Eq. \eqref{eq-ILT-appen-5}.  This expression of $\langle z_0^2(t) \rangle _c$ has been quoted in Eq. \eqref{eq-corr-eq-11} in the main text.

Let us now analyse this expression to extract the asymptotic behaviour of the variance. To perform the integration in Eq. \eqref{eq-ILT-appen-5}, we substitute the error function for large $t$ as
\begin{align}
\text{Erf}\left(\sqrt{r(t-T)} \right) \simeq 1 -\frac{e^{-r(t-T)}}{ \sqrt{\pi r (t-T)}}
\end{align}
and plug it in Eq. \eqref{eq-ILT-appen-5} as
\begin{align}
\mathcal{G}_0(t) \simeq & \frac{\mu _2(\mu _1 - \mu _2)(p+q)}{\sqrt{r}(\mu _1 - 2 \mu _2)}   \left[ \int _{0}^{t}dT~ e^{-r T} Z(T)- \frac{e^{- r t}}{\sqrt{\pi r }}~\int _{0}^{t}dT~\frac{Z(T)}{\sqrt{t-T}}\right]. \nonumber
\end{align}
For $t \to \infty$, first term inside the bracket $[..]$ simply becomes the Laplace transform $\hat{Z}(r)$ in Eq. \eqref{eq-ILT-appen-3-ps}. Therefore, one can recast $\mathcal{G}_0(t)$ as
\begin{align}
\mathcal{G}_0(t) \simeq & \frac{\mu _2(\mu _1 - \mu _2)(p+q)}{\sqrt{r}(\mu _1 - 2 \mu _2)}    \left[\hat{Z}(r)- \frac{e^{- r t}}{\sqrt{\pi r }}~\int _{0}^{t}dT~\frac{Z(T)}{\sqrt{t-T}}\right]. \label{psrs-1}
\end{align}
Next, we have to evaluate the integral over $T$ for larger values of $t$. Defining $J(t) = \int _{0}^{t}dT~\frac{Z(T)}{\sqrt{t-T}}$, we take its Laplace transform with respect to $t~(\to s)$ as
\begin{align}
\hat{J}(s) = \sqrt{\frac{\pi}{s}}~\hat{Z}(s).
\end{align}
We take small $s$ limit of this equation which corresponds to its large $t$ limit in the time domain
\begin{align}
\hat{J}(s \to 0) \simeq \frac{(\mu _1 -2 \mu _2) \sqrt{\mu _1}}{\sqrt{4(p+q)} (\mu _1-\mu _2)^2}  \sqrt{\frac{\pi}{s}}.
\end{align}
Now performing the inverse Laplace transformation, we obtain 
\begin{align}
J(t) \simeq \frac{(\mu _1 -2 \mu _2) \sqrt{\mu _1}}{\sqrt{4(p+q)t} (\mu _1-\mu _2)^2} ,~~~~\text{as }t \to \infty
\end{align}
Using this in Eq. \eqref{psrs-1}, we obtain
\begin{align}
\mathcal{G}_0(t) \simeq & \frac{\mu _1 \mu _2 (p+q)}{\sqrt{r}(\mu _1 -  \mu _2)}   \left[ \sqrt{r+4 \mu _1(p+q)}  + \frac{\mu _2 \sqrt{r}}{\mu _1 - \mu _2}\right]^{-1} - \frac{\mu _1 \sqrt{\mu _1(p+q)}}{2 r (\mu _1-\mu _2)}~\frac{e^{-rt}}{\sqrt{\pi t}}, \label{psdf-eq}
\end{align}
plugging which in Eq. \eqref{eq-12corr-eq-11} gives the relaxation behaviour of $\langle z_0^2(t)\rangle _c $ in Eqs. \eqref{eq-corr-eq-1345} and \eqref{eq-corr-eq-13}.

\section{Unequal time correlation $S_i \left(t_0, t_0+t \right)$ in the steady state $(t_0 \to \infty)$}
\label{appen-unequal}
This appendix provides a derivation of the unequal time correlation $S_i \left(t_0, t_0+t \right)$ in Eq.~\eqref{uneq-psss-eq-1}. Let us first denote by $\mathbb{S}(k,s,t)$ the joint Fourier-Laplace transform of $S_i \left(t_0, t_0+t \right)$ [see Eq.~\eqref{ghdcod-eq-2}]. We showed in Eq.~\eqref{eq:uneq_time_evln_FT} that  $\mathbb{S}(k,s,t)$ satisfies the differential equation 
\begin{align}
\frac{d \mathbb{S}(k,s,t)}{dt}=-\alpha(k) \mathbb{S}(k,s,t)+2 \pi \mu_1 (p-q) \hat{h}(s) \delta(k), \label{appen-unequal-eq-1}
\end{align}
where  $\alpha(k)=r+\mu _1(p+q)-\mu_1 p e^{j k}-\mu_1 q e^{-j k}$. Solving this equation with the initial condition $\mathbb{S}(k,s,t=0) = \mathcal{Z}(k,s)$ in Eq.~\eqref{eq-corr-eq-343} gives
\begin{align}
 \mathbb{S}(k,s,t) = \mathcal{Z}(k,s)~e^{-\alpha(k)t} + 2 \pi \hat{h}(s) h(t). \label{appen-unequal-eq-2}
\end{align}
Since we are interested in finding the correlation at the steady state (i.e. $t_0 \to \infty$), we analyse Eq.~\eqref{appen-unequal-eq-2} in the small-$s$ limit. For $\hat{h}(s)$, we use Eq.~\eqref{ashbjvie} to obtain $\hat{h}(s \to 0) \simeq \mu _1(p-q)/rs$. On the other hand, using Eq.~\eqref{eq-corr-eq-343}, we find $\mathcal{Z}(k,s)$ as
\begin{align}
\mathcal{Z}(k,s \to 0) \simeq \frac{\mathcal{B} r}{s \left[r + 2 \mu _1(p+q)(1-\cos k) \right]}  + \frac{4 \pi \mu _1^2(p-q)^2}{r^2 s}~\delta(k), \label{appen-unequal-eq-3}
\end{align}
where $\mathcal{B}$ is a constant that depends on model parameters and is given in Eq.~\eqref{uneq-psss-eq-2}. Finally plugging Eq.~\eqref{appen-unequal-eq-3} in $ \mathbb{S}(k,s,t)$ in Eq.~\eqref{appen-unequal-eq-2} and performing the inverse Fourier transformation, we find
\begin{align}
S_i \left( t_0, t_0+t\right) \Big|_{t_0 \to \infty} & =  \frac{1}{2 \pi}\int _{- \pi}^{\pi}dk~e^{-jik}~s ~\mathbb{S}(k,s,t) \Big|_{s \to 0}, \\
& =  \frac{\mathcal{B} r}{2 \pi} \int _{-\pi} ^{\pi}dk ~\frac{e^{-j i k - \alpha(k)t}}{r+2 \mu _1(p+q) (1-\cos k)} + \frac{\mu _1 ^2(p-q)^2}{r^2}(1+e^{-r t}) .\label{appen-unequal-eq-4}
\end{align}
This result has been quoted in Eq.~\eqref{uneq-psss-eq-1} in the main text.

\end{widetext}

%\bibliographystyle{apsrev}
%\bibliographystyle{jbact} 
%\bibliographystyle{dinat}
%\begin{thebibliography}{1}
\bibliography{Maindraft.bib}

\begin{thebibliography}{94}%
\makeatletter
\providecommand \@ifxundefined [1]{%
 \@ifx{#1\undefined}
}%
\providecommand \@ifnum [1]{%
 \ifnum #1\expandafter \@firstoftwo
 \else \expandafter \@secondoftwo
 \fi
}%
\providecommand \@ifx [1]{%
 \ifx #1\expandafter \@firstoftwo
 \else \expandafter \@secondoftwo
 \fi
}%
\providecommand \natexlab [1]{#1}%
\providecommand \enquote  [1]{``#1''}%
\providecommand \bibnamefont  [1]{#1}%
\providecommand \bibfnamefont [1]{#1}%
\providecommand \citenamefont [1]{#1}%
\providecommand \href@noop [0]{\@secondoftwo}%
\providecommand \href [0]{\begingroup \@sanitize@url \@href}%
\providecommand \@href[1]{\@@startlink{#1}\@@href}%
\providecommand \@@href[1]{\endgroup#1\@@endlink}%
\providecommand \@sanitize@url [0]{\catcode `\\12\catcode `\$12\catcode
  `\&12\catcode `\#12\catcode `\^12\catcode `\_12\catcode `\%12\relax}%
\providecommand \@@startlink[1]{}%
\providecommand \@@endlink[0]{}%
\providecommand \url  [0]{\begingroup\@sanitize@url \@url }%
\providecommand \@url [1]{\endgroup\@href {#1}{\urlprefix }}%
\providecommand \urlprefix  [0]{URL }%
\providecommand \Eprint [0]{\href }%
\providecommand \doibase [0]{https://doi.org/}%
\providecommand \selectlanguage [0]{\@gobble}%
\providecommand \bibinfo  [0]{\@secondoftwo}%
\providecommand \bibfield  [0]{\@secondoftwo}%
\providecommand \translation [1]{[#1]}%
\providecommand \BibitemOpen [0]{}%
\providecommand \bibitemStop [0]{}%
\providecommand \bibitemNoStop [0]{.\EOS\space}%
\providecommand \EOS [0]{\spacefactor3000\relax}%
\providecommand \BibitemShut  [1]{\csname bibitem#1\endcsname}%
\let\auto@bib@innerbib\@empty
%</preamble>
\bibitem [{\citenamefont {Lieb}\ and\ \citenamefont {Mattis}(1966)}]{Lieb1966}%
  \BibitemOpen
  \bibfield  {author} {\bibinfo {author} {\bibfnamefont {E.~H.}\ \bibnamefont
  {Lieb}}\ and\ \bibinfo {author} {\bibfnamefont {D.~C.}\ \bibnamefont
  {Mattis}},\ }\href@noop {} {\emph {\bibinfo {title} {Mathematical Physics in
  One Dimension: Exactly Solvable Models of Interacting Particles}}},\ \bibinfo
  {edition} {1st}\ ed.\ (\bibinfo  {publisher} {Academic},\ \bibinfo {year}
  {1966})\BibitemShut {NoStop}%
\bibitem [{\citenamefont {Privman}(1997)}]{privman_1997}%
  \BibitemOpen
  \bibfield  {author} {\bibinfo {author} {\bibfnamefont {V.}~\bibnamefont
  {Privman}},\ }\href {https://doi.org/10.1017/CBO9780511564284} {\emph
  {\bibinfo {title} {Nonequilibrium Statistical Mechanics in One Dimension}}}\
  (\bibinfo  {publisher} {Cambridge University Press},\ \bibinfo {year}
  {1997})\BibitemShut {NoStop}%
\bibitem [{\citenamefont {Harris}(1965)}]{harris_1965}%
  \BibitemOpen
  \bibfield  {author} {\bibinfo {author} {\bibfnamefont {T.~E.}\ \bibnamefont
  {Harris}},\ }\bibfield  {title} {\bibinfo {title} {Diffusion with collisions
  between particles},\ }\href {https://doi.org/10.2307/3212197} {\bibfield
  {journal} {\bibinfo  {journal} {Journal of Applied Probability}\ }\textbf
  {\bibinfo {volume} {2}},\ \bibinfo {pages} {323–338} (\bibinfo {year}
  {1965})}\BibitemShut {NoStop}%
\bibitem [{\citenamefont {Jepsen}(2004)}]{Jensen2004}%
  \BibitemOpen
  \bibfield  {author} {\bibinfo {author} {\bibfnamefont {D.~W.}\ \bibnamefont
  {Jepsen}},\ }\bibfield  {title} {\bibinfo {title} {{Dynamics of a Simple
  Many‐Body System of Hard Rods}},\ }\href
  {https://doi.org/10.1063/1.1704288} {\bibfield  {journal} {\bibinfo
  {journal} {Journal of Mathematical Physics}\ }\textbf {\bibinfo {volume}
  {6}},\ \bibinfo {pages} {405} (\bibinfo {year} {2004})}\BibitemShut {NoStop}%
\bibitem [{\citenamefont {Arratia}(1983)}]{10.1214/aop/1176993602}%
  \BibitemOpen
  \bibfield  {author} {\bibinfo {author} {\bibfnamefont {R.}~\bibnamefont
  {Arratia}},\ }\bibfield  {title} {\bibinfo {title} {{The Motion of a Tagged
  Particle in the Simple Symmetric Exclusion System on $Z$}},\ }\href
  {https://doi.org/10.1214/aop/1176993602} {\bibfield  {journal} {\bibinfo
  {journal} {The Annals of Probability}\ }\textbf {\bibinfo {volume} {11}},\
  \bibinfo {pages} {362 } (\bibinfo {year} {1983})}\BibitemShut {NoStop}%
\bibitem [{\citenamefont {Poncet}\ \emph {et~al.}(2021)\citenamefont {Poncet},
  \citenamefont {Grabsch}, \citenamefont {Illien},\ and\ \citenamefont
  {B\'enichou}}]{PhysRevLett.127.220601}%
  \BibitemOpen
  \bibfield  {author} {\bibinfo {author} {\bibfnamefont {A.}~\bibnamefont
  {Poncet}}, \bibinfo {author} {\bibfnamefont {A.}~\bibnamefont {Grabsch}},
  \bibinfo {author} {\bibfnamefont {P.}~\bibnamefont {Illien}},\ and\ \bibinfo
  {author} {\bibfnamefont {O.}~\bibnamefont {B\'enichou}},\ }\bibfield  {title}
  {\bibinfo {title} {Generalized correlation profiles in single-file systems},\
  }\href {https://doi.org/10.1103/PhysRevLett.127.220601} {\bibfield  {journal}
  {\bibinfo  {journal} {Phys. Rev. Lett.}\ }\textbf {\bibinfo {volume} {127}},\
  \bibinfo {pages} {220601} (\bibinfo {year} {2021})}\BibitemShut {NoStop}%
\bibitem [{\citenamefont {Grabsch}\ \emph {et~al.}(2023)\citenamefont
  {Grabsch}, \citenamefont {Rizkallah}, \citenamefont {Poncet}, \citenamefont
  {Illien},\ and\ \citenamefont {B\'enichou}}]{PhysRevE.107.044131}%
  \BibitemOpen
  \bibfield  {author} {\bibinfo {author} {\bibfnamefont {A.}~\bibnamefont
  {Grabsch}}, \bibinfo {author} {\bibfnamefont {P.}~\bibnamefont {Rizkallah}},
  \bibinfo {author} {\bibfnamefont {A.}~\bibnamefont {Poncet}}, \bibinfo
  {author} {\bibfnamefont {P.}~\bibnamefont {Illien}},\ and\ \bibinfo {author}
  {\bibfnamefont {O.}~\bibnamefont {B\'enichou}},\ }\bibfield  {title}
  {\bibinfo {title} {Exact spatial correlations in single-file diffusion},\
  }\href {https://doi.org/10.1103/PhysRevE.107.044131} {\bibfield  {journal}
  {\bibinfo  {journal} {Phys. Rev. E}\ }\textbf {\bibinfo {volume} {107}},\
  \bibinfo {pages} {044131} (\bibinfo {year} {2023})}\BibitemShut {NoStop}%
\bibitem [{\citenamefont {Krapivsky}\ \emph
  {et~al.}(2015{\natexlab{a}})\citenamefont {Krapivsky}, \citenamefont
  {Mallick},\ and\ \citenamefont {Sadhu}}]{Krapivsky_2015}%
  \BibitemOpen
  \bibfield  {author} {\bibinfo {author} {\bibfnamefont {P.~L.}\ \bibnamefont
  {Krapivsky}}, \bibinfo {author} {\bibfnamefont {K.}~\bibnamefont {Mallick}},\
  and\ \bibinfo {author} {\bibfnamefont {T.}~\bibnamefont {Sadhu}},\ }\bibfield
   {title} {\bibinfo {title} {Dynamical properties of single-file diffusion},\
  }\href {https://doi.org/10.1088/1742-5468/2015/09/P09007} {\bibfield
  {journal} {\bibinfo  {journal} {Journal of Statistical Mechanics: Theory and
  Experiment}\ }\textbf {\bibinfo {volume} {2015}},\ \bibinfo {pages} {P09007}
  (\bibinfo {year} {2015}{\natexlab{a}})}\BibitemShut {NoStop}%
\bibitem [{\citenamefont {Krapivsky}\ \emph {et~al.}(2014)\citenamefont
  {Krapivsky}, \citenamefont {Mallick},\ and\ \citenamefont
  {Sadhu}}]{SadhuPRL2014}%
  \BibitemOpen
  \bibfield  {author} {\bibinfo {author} {\bibfnamefont {P.~L.}\ \bibnamefont
  {Krapivsky}}, \bibinfo {author} {\bibfnamefont {K.}~\bibnamefont {Mallick}},\
  and\ \bibinfo {author} {\bibfnamefont {T.}~\bibnamefont {Sadhu}},\ }\bibfield
   {title} {\bibinfo {title} {Large deviations in single-file diffusion},\
  }\href {https://doi.org/10.1103/PhysRevLett.113.078101} {\bibfield  {journal}
  {\bibinfo  {journal} {Phys. Rev. Lett.}\ }\textbf {\bibinfo {volume} {113}},\
  \bibinfo {pages} {078101} (\bibinfo {year} {2014})}\BibitemShut {NoStop}%
\bibitem [{\citenamefont {Bénichou}\ \emph {et~al.}(2018)\citenamefont
  {Bénichou}, \citenamefont {Illien}, \citenamefont {Oshanin}, \citenamefont
  {Sarracino},\ and\ \citenamefont {Voituriez}}]{Benichou_2018}%
  \BibitemOpen
  \bibfield  {author} {\bibinfo {author} {\bibfnamefont {O.}~\bibnamefont
  {Bénichou}}, \bibinfo {author} {\bibfnamefont {P.}~\bibnamefont {Illien}},
  \bibinfo {author} {\bibfnamefont {G.}~\bibnamefont {Oshanin}}, \bibinfo
  {author} {\bibfnamefont {A.}~\bibnamefont {Sarracino}},\ and\ \bibinfo
  {author} {\bibfnamefont {R.}~\bibnamefont {Voituriez}},\ }\bibfield  {title}
  {\bibinfo {title} {Tracer diffusion in crowded narrow channels},\ }\href
  {https://doi.org/10.1088/1361-648X/aae13a} {\bibfield  {journal} {\bibinfo
  {journal} {Journal of Physics: Condensed Matter}\ }\textbf {\bibinfo {volume}
  {30}},\ \bibinfo {pages} {443001} (\bibinfo {year} {2018})}\BibitemShut
  {NoStop}%
\bibitem [{\citenamefont {Alexander}\ and\ \citenamefont
  {Pincus}(1978)}]{PhysRevB.18.2011}%
  \BibitemOpen
  \bibfield  {author} {\bibinfo {author} {\bibfnamefont {S.}~\bibnamefont
  {Alexander}}\ and\ \bibinfo {author} {\bibfnamefont {P.}~\bibnamefont
  {Pincus}},\ }\bibfield  {title} {\bibinfo {title} {Diffusion of labeled
  particles on one-dimensional chains},\ }\href
  {https://doi.org/10.1103/PhysRevB.18.2011} {\bibfield  {journal} {\bibinfo
  {journal} {Phys. Rev. B}\ }\textbf {\bibinfo {volume} {18}},\ \bibinfo
  {pages} {2011} (\bibinfo {year} {1978})}\BibitemShut {NoStop}%
\bibitem [{\citenamefont {Krapivsky}\ \emph
  {et~al.}(2015{\natexlab{b}})\citenamefont {Krapivsky}, \citenamefont
  {Mallick},\ and\ \citenamefont {Sadhu}}]{TSadhu2015}%
  \BibitemOpen
  \bibfield  {author} {\bibinfo {author} {\bibfnamefont {P.~L.}\ \bibnamefont
  {Krapivsky}}, \bibinfo {author} {\bibfnamefont {K.}~\bibnamefont {Mallick}},\
  and\ \bibinfo {author} {\bibfnamefont {T.}~\bibnamefont {Sadhu}},\ }\bibfield
   {title} {\bibinfo {title} {Tagged particle in single-file diffusion},\
  }\href {https://doi.org/10.1007/s10955-015-1291-0} {\bibfield  {journal}
  {\bibinfo  {journal} {Journal of Statistical Physics}\ }\textbf {\bibinfo
  {volume} {160}},\ \bibinfo {pages} {885–925} (\bibinfo {year}
  {2015}{\natexlab{b}})}\BibitemShut {NoStop}%
\bibitem [{\citenamefont {Banerjee}\ \emph
  {et~al.}(2022{\natexlab{a}})\citenamefont {Banerjee}, \citenamefont {Jack},\
  and\ \citenamefont {Cates}}]{PhysRevE.106.L062101}%
  \BibitemOpen
  \bibfield  {author} {\bibinfo {author} {\bibfnamefont {T.}~\bibnamefont
  {Banerjee}}, \bibinfo {author} {\bibfnamefont {R.~L.}\ \bibnamefont {Jack}},\
  and\ \bibinfo {author} {\bibfnamefont {M.~E.}\ \bibnamefont {Cates}},\
  }\bibfield  {title} {\bibinfo {title} {Role of initial conditions in
  one-dimensional diffusive systems: Compressibility, hyperuniformity, and
  long-term memory},\ }\href {https://doi.org/10.1103/PhysRevE.106.L062101}
  {\bibfield  {journal} {\bibinfo  {journal} {Phys. Rev. E}\ }\textbf {\bibinfo
  {volume} {106}},\ \bibinfo {pages} {L062101} (\bibinfo {year}
  {2022}{\natexlab{a}})}\BibitemShut {NoStop}%
\bibitem [{\citenamefont {Barkai}\ and\ \citenamefont
  {Silbey}(2009)}]{PhysRevLett.102.050602}%
  \BibitemOpen
  \bibfield  {author} {\bibinfo {author} {\bibfnamefont {E.}~\bibnamefont
  {Barkai}}\ and\ \bibinfo {author} {\bibfnamefont {R.}~\bibnamefont
  {Silbey}},\ }\bibfield  {title} {\bibinfo {title} {Theory of single file
  diffusion in a force field},\ }\href
  {https://doi.org/10.1103/PhysRevLett.102.050602} {\bibfield  {journal}
  {\bibinfo  {journal} {Phys. Rev. Lett.}\ }\textbf {\bibinfo {volume} {102}},\
  \bibinfo {pages} {050602} (\bibinfo {year} {2009})}\BibitemShut {NoStop}%
\bibitem [{\citenamefont {Leibovich}\ and\ \citenamefont
  {Barkai}(2013)}]{PhysRevE.88.032107}%
  \BibitemOpen
  \bibfield  {author} {\bibinfo {author} {\bibfnamefont {N.}~\bibnamefont
  {Leibovich}}\ and\ \bibinfo {author} {\bibfnamefont {E.}~\bibnamefont
  {Barkai}},\ }\bibfield  {title} {\bibinfo {title} {Everlasting effect of
  initial conditions on single-file diffusion},\ }\href
  {https://doi.org/10.1103/PhysRevE.88.032107} {\bibfield  {journal} {\bibinfo
  {journal} {Phys. Rev. E}\ }\textbf {\bibinfo {volume} {88}},\ \bibinfo
  {pages} {032107} (\bibinfo {year} {2013})}\BibitemShut {NoStop}%
\bibitem [{\citenamefont {Dandekar}\ \emph {et~al.}(2023)\citenamefont
  {Dandekar}, \citenamefont {Krapivsky},\ and\ \citenamefont
  {Mallick}}]{PhysRevE.107.044129}%
  \BibitemOpen
  \bibfield  {author} {\bibinfo {author} {\bibfnamefont {R.}~\bibnamefont
  {Dandekar}}, \bibinfo {author} {\bibfnamefont {P.~L.}\ \bibnamefont
  {Krapivsky}},\ and\ \bibinfo {author} {\bibfnamefont {K.}~\bibnamefont
  {Mallick}},\ }\bibfield  {title} {\bibinfo {title} {Dynamical fluctuations in
  the riesz gas},\ }\href {https://doi.org/10.1103/PhysRevE.107.044129}
  {\bibfield  {journal} {\bibinfo  {journal} {Phys. Rev. E}\ }\textbf {\bibinfo
  {volume} {107}},\ \bibinfo {pages} {044129} (\bibinfo {year}
  {2023})}\BibitemShut {NoStop}%
\bibitem [{\citenamefont {Hegde}\ \emph {et~al.}(2014)\citenamefont {Hegde},
  \citenamefont {Sabhapandit},\ and\ \citenamefont
  {Dhar}}]{PhysRevLett.113.120601}%
  \BibitemOpen
  \bibfield  {author} {\bibinfo {author} {\bibfnamefont {C.}~\bibnamefont
  {Hegde}}, \bibinfo {author} {\bibfnamefont {S.}~\bibnamefont {Sabhapandit}},\
  and\ \bibinfo {author} {\bibfnamefont {A.}~\bibnamefont {Dhar}},\ }\bibfield
  {title} {\bibinfo {title} {Universal large deviations for the tagged particle
  in single-file motion},\ }\href
  {https://doi.org/10.1103/PhysRevLett.113.120601} {\bibfield  {journal}
  {\bibinfo  {journal} {Phys. Rev. Lett.}\ }\textbf {\bibinfo {volume} {113}},\
  \bibinfo {pages} {120601} (\bibinfo {year} {2014})}\BibitemShut {NoStop}%
\bibitem [{\citenamefont {Roy}\ \emph {et~al.}(2013)\citenamefont {Roy},
  \citenamefont {Narayan}, \citenamefont {Dhar},\ and\ \citenamefont
  {Sabhapandit}}]{ARoy2013}%
  \BibitemOpen
  \bibfield  {author} {\bibinfo {author} {\bibfnamefont {A.}~\bibnamefont
  {Roy}}, \bibinfo {author} {\bibfnamefont {O.}~\bibnamefont {Narayan}},
  \bibinfo {author} {\bibfnamefont {A.}~\bibnamefont {Dhar}},\ and\ \bibinfo
  {author} {\bibfnamefont {S.}~\bibnamefont {Sabhapandit}},\ }\bibfield
  {title} {\bibinfo {title} {Tagged particle diffusion in one-dimensional gas
  with hamiltonian dynamics},\ }\href
  {https://doi.org/10.1007/s10955-012-0673-9} {\bibfield  {journal} {\bibinfo
  {journal} {Journal of Statistical Physics}\ }\textbf {\bibinfo {volume}
  {150}},\ \bibinfo {pages} {851–866} (\bibinfo {year} {2013})}\BibitemShut
  {NoStop}%
\bibitem [{\citenamefont {Kollmann}(2003)}]{PhysRevLett.90.180602}%
  \BibitemOpen
  \bibfield  {author} {\bibinfo {author} {\bibfnamefont {M.}~\bibnamefont
  {Kollmann}},\ }\bibfield  {title} {\bibinfo {title} {Single-file diffusion of
  atomic and colloidal systems: Asymptotic laws},\ }\href
  {https://doi.org/10.1103/PhysRevLett.90.180602} {\bibfield  {journal}
  {\bibinfo  {journal} {Phys. Rev. Lett.}\ }\textbf {\bibinfo {volume} {90}},\
  \bibinfo {pages} {180602} (\bibinfo {year} {2003})}\BibitemShut {NoStop}%
\bibitem [{\citenamefont {Burkhardt}(2019)}]{Burkhardt2019}%
  \BibitemOpen
  \bibfield  {author} {\bibinfo {author} {\bibfnamefont {T.~W.}\ \bibnamefont
  {Burkhardt}},\ }\bibfield  {title} {\bibinfo {title} {Tagged-particle
  statistics in single-file motion with random-acceleration and langevin
  dynamics},\ }\href {https://doi.org/10.1007/s10955-019-02389-y} {\bibfield
  {journal} {\bibinfo  {journal} {Journal of Statistical Physics}\ }\textbf
  {\bibinfo {volume} {177}},\ \bibinfo {pages} {806–824} (\bibinfo {year}
  {2019})}\BibitemShut {NoStop}%
\bibitem [{\citenamefont {Teomy}\ and\ \citenamefont
  {Metzler}(2019)}]{Teomy_2019}%
  \BibitemOpen
  \bibfield  {author} {\bibinfo {author} {\bibfnamefont {E.}~\bibnamefont
  {Teomy}}\ and\ \bibinfo {author} {\bibfnamefont {R.}~\bibnamefont
  {Metzler}},\ }\bibfield  {title} {\bibinfo {title} {Transport in exclusion
  processes with one-step memory: density dependence and optimal
  acceleration},\ }\href {https://doi.org/10.1088/1751-8121/ab37e4} {\bibfield
  {journal} {\bibinfo  {journal} {Journal of Physics A: Mathematical and
  Theoretical}\ }\textbf {\bibinfo {volume} {52}},\ \bibinfo {pages} {385001}
  (\bibinfo {year} {2019})}\BibitemShut {NoStop}%
\bibitem [{\citenamefont {Galanti}\ \emph {et~al.}(2013)\citenamefont
  {Galanti}, \citenamefont {Fanelli},\ and\ \citenamefont
  {Piazza}}]{Galanti2013}%
  \BibitemOpen
  \bibfield  {author} {\bibinfo {author} {\bibfnamefont {M.}~\bibnamefont
  {Galanti}}, \bibinfo {author} {\bibfnamefont {D.}~\bibnamefont {Fanelli}},\
  and\ \bibinfo {author} {\bibfnamefont {F.}~\bibnamefont {Piazza}},\
  }\bibfield  {title} {\bibinfo {title} {Persistent random walk with
  exclusion},\ }\href {https://doi.org/10.1140/epjb/e2013-40838-y} {\bibfield
  {journal} {\bibinfo  {journal} {Eur. Phys. J. B}\ }\textbf {\bibinfo {volume}
  {456}},\ \bibinfo {pages} {86} (\bibinfo {year} {2013})}\BibitemShut
  {NoStop}%
\bibitem [{\citenamefont {Dolai}\ \emph {et~al.}(2020)\citenamefont {Dolai},
  \citenamefont {Das}, \citenamefont {Kundu}, \citenamefont {Dasgupta},
  \citenamefont {Dhar},\ and\ \citenamefont {Kumar}}]{D0SM00687D}%
  \BibitemOpen
  \bibfield  {author} {\bibinfo {author} {\bibfnamefont {P.}~\bibnamefont
  {Dolai}}, \bibinfo {author} {\bibfnamefont {A.}~\bibnamefont {Das}}, \bibinfo
  {author} {\bibfnamefont {A.}~\bibnamefont {Kundu}}, \bibinfo {author}
  {\bibfnamefont {C.}~\bibnamefont {Dasgupta}}, \bibinfo {author}
  {\bibfnamefont {A.}~\bibnamefont {Dhar}},\ and\ \bibinfo {author}
  {\bibfnamefont {K.~V.}\ \bibnamefont {Kumar}},\ }\bibfield  {title} {\bibinfo
  {title} {Universal scaling in active single-file dynamics},\ }\href
  {https://doi.org/10.1039/D0SM00687D} {\bibfield  {journal} {\bibinfo
  {journal} {Soft Matter}\ }\textbf {\bibinfo {volume} {16}},\ \bibinfo {pages}
  {7077} (\bibinfo {year} {2020})}\BibitemShut {NoStop}%
\bibitem [{\citenamefont {Banerjee}\ \emph
  {et~al.}(2022{\natexlab{b}})\citenamefont {Banerjee}, \citenamefont {Jack},\
  and\ \citenamefont {Cates}}]{Banerjee_2022}%
  \BibitemOpen
  \bibfield  {author} {\bibinfo {author} {\bibfnamefont {T.}~\bibnamefont
  {Banerjee}}, \bibinfo {author} {\bibfnamefont {R.~L.}\ \bibnamefont {Jack}},\
  and\ \bibinfo {author} {\bibfnamefont {M.~E.}\ \bibnamefont {Cates}},\
  }\bibfield  {title} {\bibinfo {title} {Tracer dynamics in one dimensional
  gases of active or passive particles},\ }\href
  {https://doi.org/10.1088/1742-5468/ac4801} {\bibfield  {journal} {\bibinfo
  {journal} {Journal of Statistical Mechanics: Theory and Experiment}\ }\textbf
  {\bibinfo {volume} {2022}},\ \bibinfo {pages} {013209} (\bibinfo {year}
  {2022}{\natexlab{b}})}\BibitemShut {NoStop}%
\bibitem [{\citenamefont {Evans}\ and\ \citenamefont
  {Majumdar}(2011{\natexlab{a}})}]{PhysRevLett.106.160601}%
  \BibitemOpen
  \bibfield  {author} {\bibinfo {author} {\bibfnamefont {M.~R.}\ \bibnamefont
  {Evans}}\ and\ \bibinfo {author} {\bibfnamefont {S.~N.}\ \bibnamefont
  {Majumdar}},\ }\bibfield  {title} {\bibinfo {title} {Diffusion with
  stochastic resetting},\ }\href
  {https://doi.org/10.1103/PhysRevLett.106.160601} {\bibfield  {journal}
  {\bibinfo  {journal} {Phys. Rev. Lett.}\ }\textbf {\bibinfo {volume} {106}},\
  \bibinfo {pages} {160601} (\bibinfo {year} {2011}{\natexlab{a}})}\BibitemShut
  {NoStop}%
\bibitem [{\citenamefont {Evans}\ and\ \citenamefont
  {Majumdar}(2011{\natexlab{b}})}]{Evans_2011_1}%
  \BibitemOpen
  \bibfield  {author} {\bibinfo {author} {\bibfnamefont {M.~R.}\ \bibnamefont
  {Evans}}\ and\ \bibinfo {author} {\bibfnamefont {S.~N.}\ \bibnamefont
  {Majumdar}},\ }\bibfield  {title} {\bibinfo {title} {Diffusion with optimal
  resetting},\ }\href {https://doi.org/10.1088/1751-8113/44/43/435001}
  {\bibfield  {journal} {\bibinfo  {journal} {Journal of Physics A:
  Mathematical and Theoretical}\ }\textbf {\bibinfo {volume} {44}},\ \bibinfo
  {pages} {435001} (\bibinfo {year} {2011}{\natexlab{b}})}\BibitemShut
  {NoStop}%
\bibitem [{\citenamefont {Majumdar}\ \emph {et~al.}(2015)\citenamefont
  {Majumdar}, \citenamefont {Sabhapandit},\ and\ \citenamefont
  {Schehr}}]{PhysRevE.91.052131}%
  \BibitemOpen
  \bibfield  {author} {\bibinfo {author} {\bibfnamefont {S.~N.}\ \bibnamefont
  {Majumdar}}, \bibinfo {author} {\bibfnamefont {S.}~\bibnamefont
  {Sabhapandit}},\ and\ \bibinfo {author} {\bibfnamefont {G.}~\bibnamefont
  {Schehr}},\ }\bibfield  {title} {\bibinfo {title} {Dynamical transition in
  the temporal relaxation of stochastic processes under resetting},\ }\href
  {https://doi.org/10.1103/PhysRevE.91.052131} {\bibfield  {journal} {\bibinfo
  {journal} {Phys. Rev. E}\ }\textbf {\bibinfo {volume} {91}},\ \bibinfo
  {pages} {052131} (\bibinfo {year} {2015})}\BibitemShut {NoStop}%
\bibitem [{\citenamefont {Pal}(2015)}]{PhysRevE.91.012113}%
  \BibitemOpen
  \bibfield  {author} {\bibinfo {author} {\bibfnamefont {A.}~\bibnamefont
  {Pal}},\ }\bibfield  {title} {\bibinfo {title} {Diffusion in a potential
  landscape with stochastic resetting},\ }\href
  {https://doi.org/10.1103/PhysRevE.91.012113} {\bibfield  {journal} {\bibinfo
  {journal} {Phys. Rev. E}\ }\textbf {\bibinfo {volume} {91}},\ \bibinfo
  {pages} {012113} (\bibinfo {year} {2015})}\BibitemShut {NoStop}%
\bibitem [{\citenamefont {Pal}\ and\ \citenamefont
  {Reuveni}(2017)}]{PhysRevLett.118.030603}%
  \BibitemOpen
  \bibfield  {author} {\bibinfo {author} {\bibfnamefont {A.}~\bibnamefont
  {Pal}}\ and\ \bibinfo {author} {\bibfnamefont {S.}~\bibnamefont {Reuveni}},\
  }\bibfield  {title} {\bibinfo {title} {First passage under restart},\ }\href
  {https://doi.org/10.1103/PhysRevLett.118.030603} {\bibfield  {journal}
  {\bibinfo  {journal} {Phys. Rev. Lett.}\ }\textbf {\bibinfo {volume} {118}},\
  \bibinfo {pages} {030603} (\bibinfo {year} {2017})}\BibitemShut {NoStop}%
\bibitem [{\citenamefont {Kusmierz}\ \emph {et~al.}(2014)\citenamefont
  {Kusmierz}, \citenamefont {Majumdar}, \citenamefont {Sabhapandit},\ and\
  \citenamefont {Schehr}}]{PhysRevLett.113.220602}%
  \BibitemOpen
  \bibfield  {author} {\bibinfo {author} {\bibfnamefont {L.}~\bibnamefont
  {Kusmierz}}, \bibinfo {author} {\bibfnamefont {S.~N.}\ \bibnamefont
  {Majumdar}}, \bibinfo {author} {\bibfnamefont {S.}~\bibnamefont
  {Sabhapandit}},\ and\ \bibinfo {author} {\bibfnamefont {G.}~\bibnamefont
  {Schehr}},\ }\bibfield  {title} {\bibinfo {title} {First order transition for
  the optimal search time of l\'evy flights with resetting},\ }\href
  {https://doi.org/10.1103/PhysRevLett.113.220602} {\bibfield  {journal}
  {\bibinfo  {journal} {Phys. Rev. Lett.}\ }\textbf {\bibinfo {volume} {113}},\
  \bibinfo {pages} {220602} (\bibinfo {year} {2014})}\BibitemShut {NoStop}%
\bibitem [{\citenamefont {Bressloff}(2020{\natexlab{a}})}]{Bressloff_2020gfs}%
  \BibitemOpen
  \bibfield  {author} {\bibinfo {author} {\bibfnamefont {P.~C.}\ \bibnamefont
  {Bressloff}},\ }\bibfield  {title} {\bibinfo {title} {Directed intermittent
  search with stochastic resetting},\ }\href
  {https://doi.org/10.1088/1751-8121/ab7138} {\bibfield  {journal} {\bibinfo
  {journal} {Journal of Physics A: Mathematical and Theoretical}\ }\textbf
  {\bibinfo {volume} {53}},\ \bibinfo {pages} {105001} (\bibinfo {year}
  {2020}{\natexlab{a}})}\BibitemShut {NoStop}%
\bibitem [{\citenamefont {Reuveni}(2016)}]{PhysRevLett.116.170601}%
  \BibitemOpen
  \bibfield  {author} {\bibinfo {author} {\bibfnamefont {S.}~\bibnamefont
  {Reuveni}},\ }\bibfield  {title} {\bibinfo {title} {Optimal stochastic
  restart renders fluctuations in first passage times universal},\ }\href
  {https://doi.org/10.1103/PhysRevLett.116.170601} {\bibfield  {journal}
  {\bibinfo  {journal} {Phys. Rev. Lett.}\ }\textbf {\bibinfo {volume} {116}},\
  \bibinfo {pages} {170601} (\bibinfo {year} {2016})}\BibitemShut {NoStop}%
\bibitem [{\citenamefont {Gupta}(2019)}]{Gupta_2019cjd}%
  \BibitemOpen
  \bibfield  {author} {\bibinfo {author} {\bibfnamefont {D.}~\bibnamefont
  {Gupta}},\ }\bibfield  {title} {\bibinfo {title} {Stochastic resetting in
  underdamped brownian motion},\ }\href
  {https://doi.org/10.1088/1742-5468/ab054a} {\bibfield  {journal} {\bibinfo
  {journal} {Journal of Statistical Mechanics: Theory and Experiment}\ }\textbf
  {\bibinfo {volume} {2019}},\ \bibinfo {pages} {033212} (\bibinfo {year}
  {2019})}\BibitemShut {NoStop}%
\bibitem [{\citenamefont {Singh}(2020)}]{Singh_20202d}%
  \BibitemOpen
  \bibfield  {author} {\bibinfo {author} {\bibfnamefont {P.}~\bibnamefont
  {Singh}},\ }\bibfield  {title} {\bibinfo {title} {Random acceleration process
  under stochastic resetting},\ }\href
  {https://doi.org/10.1088/1751-8121/abaf2d} {\bibfield  {journal} {\bibinfo
  {journal} {Journal of Physics A: Mathematical and Theoretical}\ }\textbf
  {\bibinfo {volume} {53}},\ \bibinfo {pages} {405005} (\bibinfo {year}
  {2020})}\BibitemShut {NoStop}%
\bibitem [{\citenamefont {Evans}\ and\ \citenamefont
  {Majumdar}(2018)}]{Evans_2018vaf}%
  \BibitemOpen
  \bibfield  {author} {\bibinfo {author} {\bibfnamefont {M.~R.}\ \bibnamefont
  {Evans}}\ and\ \bibinfo {author} {\bibfnamefont {S.~N.}\ \bibnamefont
  {Majumdar}},\ }\bibfield  {title} {\bibinfo {title} {Run and tumble particle
  under resetting: a renewal approach},\ }\href
  {https://doi.org/10.1088/1751-8121/aae74e} {\bibfield  {journal} {\bibinfo
  {journal} {Journal of Physics A: Mathematical and Theoretical}\ }\textbf
  {\bibinfo {volume} {51}},\ \bibinfo {pages} {475003} (\bibinfo {year}
  {2018})}\BibitemShut {NoStop}%
\bibitem [{\citenamefont {Kumar}\ \emph {et~al.}(2020)\citenamefont {Kumar},
  \citenamefont {Sadekar},\ and\ \citenamefont {Basu}}]{PhysRevE.102.052129}%
  \BibitemOpen
  \bibfield  {author} {\bibinfo {author} {\bibfnamefont {V.}~\bibnamefont
  {Kumar}}, \bibinfo {author} {\bibfnamefont {O.}~\bibnamefont {Sadekar}},\
  and\ \bibinfo {author} {\bibfnamefont {U.}~\bibnamefont {Basu}},\ }\bibfield
  {title} {\bibinfo {title} {Active brownian motion in two dimensions under
  stochastic resetting},\ }\href {https://doi.org/10.1103/PhysRevE.102.052129}
  {\bibfield  {journal} {\bibinfo  {journal} {Phys. Rev. E}\ }\textbf {\bibinfo
  {volume} {102}},\ \bibinfo {pages} {052129} (\bibinfo {year}
  {2020})}\BibitemShut {NoStop}%
\bibitem [{\citenamefont {Singh}\ and\ \citenamefont
  {Pal}(2021)}]{PhysRevE.103.052119}%
  \BibitemOpen
  \bibfield  {author} {\bibinfo {author} {\bibfnamefont {P.}~\bibnamefont
  {Singh}}\ and\ \bibinfo {author} {\bibfnamefont {A.}~\bibnamefont {Pal}},\
  }\bibfield  {title} {\bibinfo {title} {Extremal statistics for stochastic
  resetting systems},\ }\href {https://doi.org/10.1103/PhysRevE.103.052119}
  {\bibfield  {journal} {\bibinfo  {journal} {Phys. Rev. E}\ }\textbf {\bibinfo
  {volume} {103}},\ \bibinfo {pages} {052119} (\bibinfo {year}
  {2021})}\BibitemShut {NoStop}%
\bibitem [{\citenamefont {Stojkoski}\ \emph
  {et~al.}(2022{\natexlab{a}})\citenamefont {Stojkoski}, \citenamefont
  {Sandev}, \citenamefont {Kocarev},\ and\ \citenamefont
  {Pal}}]{Stojkoski_2022_auto}%
  \BibitemOpen
  \bibfield  {author} {\bibinfo {author} {\bibfnamefont {V.}~\bibnamefont
  {Stojkoski}}, \bibinfo {author} {\bibfnamefont {T.}~\bibnamefont {Sandev}},
  \bibinfo {author} {\bibfnamefont {L.}~\bibnamefont {Kocarev}},\ and\ \bibinfo
  {author} {\bibfnamefont {A.}~\bibnamefont {Pal}},\ }\bibfield  {title}
  {\bibinfo {title} {Autocorrelation functions and ergodicity in diffusion with
  stochastic resetting},\ }\href {https://doi.org/10.1088/1751-8121/ac4ce9}
  {\bibfield  {journal} {\bibinfo  {journal} {Journal of Physics A:
  Mathematical and Theoretical}\ }\textbf {\bibinfo {volume} {55}},\ \bibinfo
  {pages} {104003} (\bibinfo {year} {2022}{\natexlab{a}})}\BibitemShut
  {NoStop}%
\bibitem [{\citenamefont {Majumdar}\ and\ \citenamefont
  {Oshanin}(2018)}]{Majumdar_2018_auto}%
  \BibitemOpen
  \bibfield  {author} {\bibinfo {author} {\bibfnamefont {S.~N.}\ \bibnamefont
  {Majumdar}}\ and\ \bibinfo {author} {\bibfnamefont {G.}~\bibnamefont
  {Oshanin}},\ }\bibfield  {title} {\bibinfo {title} {Spectral content of
  fractional brownian motion with stochastic reset},\ }\href
  {https://doi.org/10.1088/1751-8121/aadef0} {\bibfield  {journal} {\bibinfo
  {journal} {Journal of Physics A: Mathematical and Theoretical}\ }\textbf
  {\bibinfo {volume} {51}},\ \bibinfo {pages} {435001} (\bibinfo {year}
  {2018})}\BibitemShut {NoStop}%
\bibitem [{\citenamefont {Singh}\ and\ \citenamefont
  {Pal}(2022)}]{Singh_2022aphnx}%
  \BibitemOpen
  \bibfield  {author} {\bibinfo {author} {\bibfnamefont {P.}~\bibnamefont
  {Singh}}\ and\ \bibinfo {author} {\bibfnamefont {A.}~\bibnamefont {Pal}},\
  }\bibfield  {title} {\bibinfo {title} {First-passage brownian functionals
  with stochastic resetting},\ }\href
  {https://doi.org/10.1088/1751-8121/ac677c} {\bibfield  {journal} {\bibinfo
  {journal} {Journal of Physics A: Mathematical and Theoretical}\ }\textbf
  {\bibinfo {volume} {55}},\ \bibinfo {pages} {234001} (\bibinfo {year}
  {2022})}\BibitemShut {NoStop}%
\bibitem [{\citenamefont {Pal}\ \emph {et~al.}(2022)\citenamefont {Pal},
  \citenamefont {Kostinski},\ and\ \citenamefont {Reuveni}}]{Pal_2022jafda12}%
  \BibitemOpen
  \bibfield  {author} {\bibinfo {author} {\bibfnamefont {A.}~\bibnamefont
  {Pal}}, \bibinfo {author} {\bibfnamefont {S.}~\bibnamefont {Kostinski}},\
  and\ \bibinfo {author} {\bibfnamefont {S.}~\bibnamefont {Reuveni}},\
  }\bibfield  {title} {\bibinfo {title} {The inspection paradox in stochastic
  resetting},\ }\href {https://doi.org/10.1088/1751-8121/ac3cdf} {\bibfield
  {journal} {\bibinfo  {journal} {Journal of Physics A: Mathematical and
  Theoretical}\ }\textbf {\bibinfo {volume} {55}},\ \bibinfo {pages} {021001}
  (\bibinfo {year} {2022})}\BibitemShut {NoStop}%
\bibitem [{\citenamefont {Pal}\ \emph {et~al.}(2020)\citenamefont {Pal},
  \citenamefont {Ku\ifmmode~\acute{s}\else \'{s}\fi{}mierz},\ and\
  \citenamefont {Reuveni}}]{PhysRevResearch.2.043174}%
  \BibitemOpen
  \bibfield  {author} {\bibinfo {author} {\bibfnamefont {A.}~\bibnamefont
  {Pal}}, \bibinfo {author} {\bibfnamefont {L.}~\bibnamefont
  {Ku\ifmmode~\acute{s}\else \'{s}\fi{}mierz}},\ and\ \bibinfo {author}
  {\bibfnamefont {S.}~\bibnamefont {Reuveni}},\ }\bibfield  {title} {\bibinfo
  {title} {Search with home returns provides advantage under high
  uncertainty},\ }\href {https://doi.org/10.1103/PhysRevResearch.2.043174}
  {\bibfield  {journal} {\bibinfo  {journal} {Phys. Rev. Res.}\ }\textbf
  {\bibinfo {volume} {2}},\ \bibinfo {pages} {043174} (\bibinfo {year}
  {2020})}\BibitemShut {NoStop}%
\bibitem [{\citenamefont {Montanari}\ and\ \citenamefont
  {Zecchina}(2002)}]{PhysRevLett.88.178701}%
  \BibitemOpen
  \bibfield  {author} {\bibinfo {author} {\bibfnamefont {A.}~\bibnamefont
  {Montanari}}\ and\ \bibinfo {author} {\bibfnamefont {R.}~\bibnamefont
  {Zecchina}},\ }\bibfield  {title} {\bibinfo {title} {Optimizing searches via
  rare events},\ }\href {https://doi.org/10.1103/PhysRevLett.88.178701}
  {\bibfield  {journal} {\bibinfo  {journal} {Phys. Rev. Lett.}\ }\textbf
  {\bibinfo {volume} {88}},\ \bibinfo {pages} {178701} (\bibinfo {year}
  {2002})}\BibitemShut {NoStop}%
\bibitem [{\citenamefont {Luby}\ \emph {et~al.}(1993)\citenamefont {Luby},
  \citenamefont {Sinclair},\ and\ \citenamefont {Zuckerman}}]{LUBY1993173}%
  \BibitemOpen
  \bibfield  {author} {\bibinfo {author} {\bibfnamefont {M.}~\bibnamefont
  {Luby}}, \bibinfo {author} {\bibfnamefont {A.}~\bibnamefont {Sinclair}},\
  and\ \bibinfo {author} {\bibfnamefont {D.}~\bibnamefont {Zuckerman}},\
  }\bibfield  {title} {\bibinfo {title} {Optimal speedup of las vegas
  algorithms},\ }\href
  {https://doi.org/https://doi.org/10.1016/0020-0190(93)90029-9} {\bibfield
  {journal} {\bibinfo  {journal} {Information Processing Letters}\ }\textbf
  {\bibinfo {volume} {47}},\ \bibinfo {pages} {173} (\bibinfo {year}
  {1993})}\BibitemShut {NoStop}%
\bibitem [{\citenamefont {Hamlin}\ \emph {et~al.}(2019)\citenamefont {Hamlin},
  \citenamefont {Thrasher}, \citenamefont {Keyrouz},\ and\ \citenamefont
  {Mascagni}}]{HamlinThrasherKeyrouzMascagni+2019+329+340}%
  \BibitemOpen
  \bibfield  {author} {\bibinfo {author} {\bibfnamefont {P.}~\bibnamefont
  {Hamlin}}, \bibinfo {author} {\bibfnamefont {W.~J.}\ \bibnamefont
  {Thrasher}}, \bibinfo {author} {\bibfnamefont {W.}~\bibnamefont {Keyrouz}},\
  and\ \bibinfo {author} {\bibfnamefont {M.}~\bibnamefont {Mascagni}},\
  }\bibfield  {title} {\bibinfo {title} {Geometry entrapment in
  walk-on-subdomains},\ }\href {https://doi.org/doi:10.1515/mcma-2019-2052}
  {\bibfield  {journal} {\bibinfo  {journal} {Monte Carlo Methods and
  Applications}\ }\textbf {\bibinfo {volume} {25}},\ \bibinfo {pages} {329}
  (\bibinfo {year} {2019})}\BibitemShut {NoStop}%
\bibitem [{\citenamefont {Reuveni}\ \emph {et~al.}(2014)\citenamefont
  {Reuveni}, \citenamefont {Urbakh},\ and\ \citenamefont
  {Klafter}}]{doi:10.1073/pnas.1318122111}%
  \BibitemOpen
  \bibfield  {author} {\bibinfo {author} {\bibfnamefont {S.}~\bibnamefont
  {Reuveni}}, \bibinfo {author} {\bibfnamefont {M.}~\bibnamefont {Urbakh}},\
  and\ \bibinfo {author} {\bibfnamefont {J.}~\bibnamefont {Klafter}},\
  }\bibfield  {title} {\bibinfo {title} {Role of substrate unbinding in
  michaelis–menten enzymatic reactions},\ }\href
  {https://doi.org/10.1073/pnas.1318122111} {\bibfield  {journal} {\bibinfo
  {journal} {Proceedings of the National Academy of Sciences}\ }\textbf
  {\bibinfo {volume} {111}},\ \bibinfo {pages} {4391} (\bibinfo {year}
  {2014})},\ \Eprint
  {https://arxiv.org/abs/https://www.pnas.org/doi/pdf/10.1073/pnas.1318122111}
  {https://www.pnas.org/doi/pdf/10.1073/pnas.1318122111} \BibitemShut {NoStop}%
\bibitem [{\citenamefont {Rold\'an}\ \emph {et~al.}(2016)\citenamefont
  {Rold\'an}, \citenamefont {Lisica}, \citenamefont {S\'anchez-Taltavull},\
  and\ \citenamefont {Grill}}]{PhysRevE.93.062411}%
  \BibitemOpen
  \bibfield  {author} {\bibinfo {author} {\bibfnamefont {E.}~\bibnamefont
  {Rold\'an}}, \bibinfo {author} {\bibfnamefont {A.}~\bibnamefont {Lisica}},
  \bibinfo {author} {\bibfnamefont {D.}~\bibnamefont {S\'anchez-Taltavull}},\
  and\ \bibinfo {author} {\bibfnamefont {S.~W.}\ \bibnamefont {Grill}},\
  }\bibfield  {title} {\bibinfo {title} {Stochastic resetting in backtrack
  recovery by rna polymerases},\ }\href
  {https://doi.org/10.1103/PhysRevE.93.062411} {\bibfield  {journal} {\bibinfo
  {journal} {Phys. Rev. E}\ }\textbf {\bibinfo {volume} {93}},\ \bibinfo
  {pages} {062411} (\bibinfo {year} {2016})}\BibitemShut {NoStop}%
\bibitem [{\citenamefont {Bressloff}(2020{\natexlab{b}})}]{PaulBressloff_2020}%
  \BibitemOpen
  \bibfield  {author} {\bibinfo {author} {\bibfnamefont {P.~C.}\ \bibnamefont
  {Bressloff}},\ }\bibfield  {title} {\bibinfo {title} {Modeling active
  cellular transport as a directed search process with stochastic resetting and
  delays},\ }\href {https://doi.org/10.1088/1751-8121/ab9fb7} {\bibfield
  {journal} {\bibinfo  {journal} {Journal of Physics A: Mathematical and
  Theoretical}\ }\textbf {\bibinfo {volume} {53}},\ \bibinfo {pages} {355001}
  (\bibinfo {year} {2020}{\natexlab{b}})}\BibitemShut {NoStop}%
\bibitem [{\citenamefont {Pal}\ \emph {et~al.}(2021)\citenamefont {Pal},
  \citenamefont {Reuveni},\ and\ \citenamefont
  {Rahav}}]{PhysRevResearch.3.L032034}%
  \BibitemOpen
  \bibfield  {author} {\bibinfo {author} {\bibfnamefont {A.}~\bibnamefont
  {Pal}}, \bibinfo {author} {\bibfnamefont {S.}~\bibnamefont {Reuveni}},\ and\
  \bibinfo {author} {\bibfnamefont {S.}~\bibnamefont {Rahav}},\ }\bibfield
  {title} {\bibinfo {title} {Thermodynamic uncertainty relation for
  first-passage times on markov chains},\ }\href
  {https://doi.org/10.1103/PhysRevResearch.3.L032034} {\bibfield  {journal}
  {\bibinfo  {journal} {Phys. Rev. Res.}\ }\textbf {\bibinfo {volume} {3}},\
  \bibinfo {pages} {L032034} (\bibinfo {year} {2021})}\BibitemShut {NoStop}%
\bibitem [{\citenamefont {Nagar}\ and\ \citenamefont
  {Gupta}(2016)}]{PhysRevE.93.060102}%
  \BibitemOpen
  \bibfield  {author} {\bibinfo {author} {\bibfnamefont {A.}~\bibnamefont
  {Nagar}}\ and\ \bibinfo {author} {\bibfnamefont {S.}~\bibnamefont {Gupta}},\
  }\bibfield  {title} {\bibinfo {title} {Diffusion with stochastic resetting at
  power-law times},\ }\href {https://doi.org/10.1103/PhysRevE.93.060102}
  {\bibfield  {journal} {\bibinfo  {journal} {Phys. Rev. E}\ }\textbf {\bibinfo
  {volume} {93}},\ \bibinfo {pages} {060102} (\bibinfo {year}
  {2016})}\BibitemShut {NoStop}%
\bibitem [{\citenamefont {Chechkin}\ and\ \citenamefont
  {Sokolov}(2018)}]{PhysRevLett.121.050601}%
  \BibitemOpen
  \bibfield  {author} {\bibinfo {author} {\bibfnamefont {A.}~\bibnamefont
  {Chechkin}}\ and\ \bibinfo {author} {\bibfnamefont {I.~M.}\ \bibnamefont
  {Sokolov}},\ }\bibfield  {title} {\bibinfo {title} {Random search with
  resetting: A unified renewal approach},\ }\href
  {https://doi.org/10.1103/PhysRevLett.121.050601} {\bibfield  {journal}
  {\bibinfo  {journal} {Phys. Rev. Lett.}\ }\textbf {\bibinfo {volume} {121}},\
  \bibinfo {pages} {050601} (\bibinfo {year} {2018})}\BibitemShut {NoStop}%
\bibitem [{\citenamefont {Mukherjee}\ \emph {et~al.}(2018)\citenamefont
  {Mukherjee}, \citenamefont {Sengupta},\ and\ \citenamefont
  {Majumdar}}]{PhysRevB.98.104309}%
  \BibitemOpen
  \bibfield  {author} {\bibinfo {author} {\bibfnamefont {B.}~\bibnamefont
  {Mukherjee}}, \bibinfo {author} {\bibfnamefont {K.}~\bibnamefont
  {Sengupta}},\ and\ \bibinfo {author} {\bibfnamefont {S.~N.}\ \bibnamefont
  {Majumdar}},\ }\bibfield  {title} {\bibinfo {title} {Quantum dynamics with
  stochastic reset},\ }\href {https://doi.org/10.1103/PhysRevB.98.104309}
  {\bibfield  {journal} {\bibinfo  {journal} {Phys. Rev. B}\ }\textbf {\bibinfo
  {volume} {98}},\ \bibinfo {pages} {104309} (\bibinfo {year}
  {2018})}\BibitemShut {NoStop}%
\bibitem [{\citenamefont {Kulkarni}\ and\ \citenamefont
  {Majumdar}(2023)}]{Kulkarni2023}%
  \BibitemOpen
  \bibfield  {author} {\bibinfo {author} {\bibfnamefont {M.}~\bibnamefont
  {Kulkarni}}\ and\ \bibinfo {author} {\bibfnamefont {S.~N.}\ \bibnamefont
  {Majumdar}},\ }\href@noop {} {\bibinfo {title} {Generating entanglement by
  quantum resetting}},\ \bibinfo {howpublished}
  {\url{https://arxiv.org/abs/2307.07485}} (\bibinfo {year} {2023}),\ \bibinfo
  {note} {arXiv:2307.07485}\BibitemShut {NoStop}%
\bibitem [{\citenamefont {Yin}\ and\ \citenamefont
  {Barkai}(2023{\natexlab{a}})}]{PhysRevLett.130.050802}%
  \BibitemOpen
  \bibfield  {author} {\bibinfo {author} {\bibfnamefont {R.}~\bibnamefont
  {Yin}}\ and\ \bibinfo {author} {\bibfnamefont {E.}~\bibnamefont {Barkai}},\
  }\bibfield  {title} {\bibinfo {title} {Restart expedites quantum walk hitting
  times},\ }\href {https://doi.org/10.1103/PhysRevLett.130.050802} {\bibfield
  {journal} {\bibinfo  {journal} {Phys. Rev. Lett.}\ }\textbf {\bibinfo
  {volume} {130}},\ \bibinfo {pages} {050802} (\bibinfo {year}
  {2023}{\natexlab{a}})}\BibitemShut {NoStop}%
\bibitem [{\citenamefont {Yin}\ and\ \citenamefont
  {Barkai}(2023{\natexlab{b}})}]{Yin2023}%
  \BibitemOpen
  \bibfield  {author} {\bibinfo {author} {\bibfnamefont {R.}~\bibnamefont
  {Yin}}\ and\ \bibinfo {author} {\bibfnamefont {E.}~\bibnamefont {Barkai}},\
  }\href@noop {} {\bibinfo {title} {Instability in the quantum restart
  problem}},\ \bibinfo {howpublished} {\url{https://arxiv.org/abs/2301.06100}}
  (\bibinfo {year} {2023}{\natexlab{b}}),\ \bibinfo {note}
  {arXiv:2301.06100}\BibitemShut {NoStop}%
\bibitem [{\citenamefont {Dubey}\ \emph {et~al.}(2023)\citenamefont {Dubey},
  \citenamefont {Chetrite},\ and\ \citenamefont {Dhar}}]{Dubey_2023}%
  \BibitemOpen
  \bibfield  {author} {\bibinfo {author} {\bibfnamefont {V.}~\bibnamefont
  {Dubey}}, \bibinfo {author} {\bibfnamefont {R.}~\bibnamefont {Chetrite}},\
  and\ \bibinfo {author} {\bibfnamefont {A.}~\bibnamefont {Dhar}},\ }\bibfield
  {title} {\bibinfo {title} {Quantum resetting in continuous measurement
  induced dynamics of a qubit},\ }\href
  {https://doi.org/10.1088/1751-8121/acc290} {\bibfield  {journal} {\bibinfo
  {journal} {Journal of Physics A: Mathematical and Theoretical}\ }\textbf
  {\bibinfo {volume} {56}},\ \bibinfo {pages} {154001} (\bibinfo {year}
  {2023})}\BibitemShut {NoStop}%
\bibitem [{\citenamefont {Tal-Friedman}\ \emph {et~al.}(2020)\citenamefont
  {Tal-Friedman}, \citenamefont {Pal}, \citenamefont {Sekhon}, \citenamefont
  {Reuveni},\ and\ \citenamefont {Roichman}}]{reset-Exp1}%
  \BibitemOpen
  \bibfield  {author} {\bibinfo {author} {\bibfnamefont {O.}~\bibnamefont
  {Tal-Friedman}}, \bibinfo {author} {\bibfnamefont {A.}~\bibnamefont {Pal}},
  \bibinfo {author} {\bibfnamefont {A.}~\bibnamefont {Sekhon}}, \bibinfo
  {author} {\bibfnamefont {S.}~\bibnamefont {Reuveni}},\ and\ \bibinfo {author}
  {\bibfnamefont {Y.}~\bibnamefont {Roichman}},\ }\bibfield  {title} {\bibinfo
  {title} {Experimental realization of diffusion with stochastic resetting},\
  }\href {https://doi.org/10.1021/acs.jpclett.0c02122} {\bibfield  {journal}
  {\bibinfo  {journal} {The Journal of Physical Chemistry Letters}\ }\textbf
  {\bibinfo {volume} {11}},\ \bibinfo {pages} {7350} (\bibinfo {year}
  {2020})},\ \bibinfo {note} {pMID: 32787296},\ \Eprint
  {https://arxiv.org/abs/https://doi.org/10.1021/acs.jpclett.0c02122}
  {https://doi.org/10.1021/acs.jpclett.0c02122} \BibitemShut {NoStop}%
\bibitem [{\citenamefont {Besga}\ \emph {et~al.}(2020)\citenamefont {Besga},
  \citenamefont {Bovon}, \citenamefont {Petrosyan}, \citenamefont {Majumdar},\
  and\ \citenamefont {Ciliberto}}]{reset-Exp2}%
  \BibitemOpen
  \bibfield  {author} {\bibinfo {author} {\bibfnamefont {B.}~\bibnamefont
  {Besga}}, \bibinfo {author} {\bibfnamefont {A.}~\bibnamefont {Bovon}},
  \bibinfo {author} {\bibfnamefont {A.}~\bibnamefont {Petrosyan}}, \bibinfo
  {author} {\bibfnamefont {S.~N.}\ \bibnamefont {Majumdar}},\ and\ \bibinfo
  {author} {\bibfnamefont {S.}~\bibnamefont {Ciliberto}},\ }\bibfield  {title}
  {\bibinfo {title} {Optimal mean first-passage time for a brownian searcher
  subjected to resetting: Experimental and theoretical results},\ }\href
  {https://doi.org/10.1103/PhysRevResearch.2.032029} {\bibfield  {journal}
  {\bibinfo  {journal} {Phys. Rev. Res.}\ }\textbf {\bibinfo {volume} {2}},\
  \bibinfo {pages} {032029} (\bibinfo {year} {2020})}\BibitemShut {NoStop}%
\bibitem [{\citenamefont {Besga}\ \emph {et~al.}(2021)\citenamefont {Besga},
  \citenamefont {Faisant}, \citenamefont {Petrosyan}, \citenamefont
  {Ciliberto},\ and\ \citenamefont {Majumdar}}]{reset-Exp3}%
  \BibitemOpen
  \bibfield  {author} {\bibinfo {author} {\bibfnamefont {B.}~\bibnamefont
  {Besga}}, \bibinfo {author} {\bibfnamefont {F.}~\bibnamefont {Faisant}},
  \bibinfo {author} {\bibfnamefont {A.}~\bibnamefont {Petrosyan}}, \bibinfo
  {author} {\bibfnamefont {S.}~\bibnamefont {Ciliberto}},\ and\ \bibinfo
  {author} {\bibfnamefont {S.~N.}\ \bibnamefont {Majumdar}},\ }\bibfield
  {title} {\bibinfo {title} {Dynamical phase transition in the first-passage
  probability of a brownian motion},\ }\href
  {https://doi.org/10.1103/PhysRevE.104.L012102} {\bibfield  {journal}
  {\bibinfo  {journal} {Phys. Rev. E}\ }\textbf {\bibinfo {volume} {104}},\
  \bibinfo {pages} {L012102} (\bibinfo {year} {2021})}\BibitemShut {NoStop}%
\bibitem [{\citenamefont {Evans}\ \emph {et~al.}(2020)\citenamefont {Evans},
  \citenamefont {Majumdar},\ and\ \citenamefont {Schehr}}]{reset-review1}%
  \BibitemOpen
  \bibfield  {author} {\bibinfo {author} {\bibfnamefont {M.~R.}\ \bibnamefont
  {Evans}}, \bibinfo {author} {\bibfnamefont {S.~N.}\ \bibnamefont
  {Majumdar}},\ and\ \bibinfo {author} {\bibfnamefont {G.}~\bibnamefont
  {Schehr}},\ }\bibfield  {title} {\bibinfo {title} {Stochastic resetting and
  applications},\ }\href {https://doi.org/10.1088/1751-8121/ab7cfe} {\bibfield
  {journal} {\bibinfo  {journal} {Journal of Physics A: Mathematical and
  Theoretical}\ }\textbf {\bibinfo {volume} {53}},\ \bibinfo {pages} {193001}
  (\bibinfo {year} {2020})}\BibitemShut {NoStop}%
\bibitem [{\citenamefont {Gupta}\ and\ \citenamefont
  {Jayannavar}(2022)}]{reset-review2}%
  \BibitemOpen
  \bibfield  {author} {\bibinfo {author} {\bibfnamefont {S.}~\bibnamefont
  {Gupta}}\ and\ \bibinfo {author} {\bibfnamefont {A.~M.}\ \bibnamefont
  {Jayannavar}},\ }\bibfield  {title} {\bibinfo {title} {Stochastic resetting:
  A (very) brief review},\ }\bibfield  {journal} {\bibinfo  {journal}
  {Frontiers in Physics}\ }\textbf {\bibinfo {volume} {10}},\ \href
  {https://doi.org/10.3389/fphy.2022.789097} {10.3389/fphy.2022.789097}
  (\bibinfo {year} {2022})\BibitemShut {NoStop}%
\bibitem [{\citenamefont {Nagar}\ and\ \citenamefont
  {Gupta}(2023)}]{reset-review3}%
  \BibitemOpen
  \bibfield  {author} {\bibinfo {author} {\bibfnamefont {A.}~\bibnamefont
  {Nagar}}\ and\ \bibinfo {author} {\bibfnamefont {S.}~\bibnamefont {Gupta}},\
  }\bibfield  {title} {\bibinfo {title} {Stochastic resetting in interacting
  particle systems: a review},\ }\href
  {https://doi.org/10.1088/1751-8121/acda6c} {\bibfield  {journal} {\bibinfo
  {journal} {Journal of Physics A: Mathematical and Theoretical}\ }\textbf
  {\bibinfo {volume} {56}},\ \bibinfo {pages} {283001} (\bibinfo {year}
  {2023})}\BibitemShut {NoStop}%
\bibitem [{\citenamefont {Pal}\ \emph {et~al.}(2023)\citenamefont {Pal},
  \citenamefont {Stojkoski},\ and\ \citenamefont {Sandev}}]{reset-review4}%
  \BibitemOpen
  \bibfield  {author} {\bibinfo {author} {\bibfnamefont {A.}~\bibnamefont
  {Pal}}, \bibinfo {author} {\bibfnamefont {V.}~\bibnamefont {Stojkoski}},\
  and\ \bibinfo {author} {\bibfnamefont {T.}~\bibnamefont {Sandev}},\
  }\href@noop {} {\bibinfo {title} {Random resetting in search problems}},\
  \bibinfo {howpublished} {\url{https://arxiv.org/abs/2310.12057}} (\bibinfo
  {year} {2023}),\ \bibinfo {note} {arXiv:2310.12057}\BibitemShut {NoStop}%
\bibitem [{\citenamefont {Basu}\ \emph {et~al.}(2019)\citenamefont {Basu},
  \citenamefont {Kundu},\ and\ \citenamefont {Pal}}]{PhysRevE.100.032136}%
  \BibitemOpen
  \bibfield  {author} {\bibinfo {author} {\bibfnamefont {U.}~\bibnamefont
  {Basu}}, \bibinfo {author} {\bibfnamefont {A.}~\bibnamefont {Kundu}},\ and\
  \bibinfo {author} {\bibfnamefont {A.}~\bibnamefont {Pal}},\ }\bibfield
  {title} {\bibinfo {title} {Symmetric exclusion process under stochastic
  resetting},\ }\href {https://doi.org/10.1103/PhysRevE.100.032136} {\bibfield
  {journal} {\bibinfo  {journal} {Phys. Rev. E}\ }\textbf {\bibinfo {volume}
  {100}},\ \bibinfo {pages} {032136} (\bibinfo {year} {2019})}\BibitemShut
  {NoStop}%
\bibitem [{\citenamefont {Sadekar}\ and\ \citenamefont
  {Basu}(2020)}]{Sadekar_2020-reset}%
  \BibitemOpen
  \bibfield  {author} {\bibinfo {author} {\bibfnamefont {O.}~\bibnamefont
  {Sadekar}}\ and\ \bibinfo {author} {\bibfnamefont {U.}~\bibnamefont {Basu}},\
  }\bibfield  {title} {\bibinfo {title} {Zero-current nonequilibrium state in
  symmetric exclusion process with dichotomous stochastic resetting},\ }\href
  {https://doi.org/10.1088/1742-5468/ab9e5e} {\bibfield  {journal} {\bibinfo
  {journal} {Journal of Statistical Mechanics: Theory and Experiment}\ }\textbf
  {\bibinfo {volume} {2020}},\ \bibinfo {pages} {073209} (\bibinfo {year}
  {2020})}\BibitemShut {NoStop}%
\bibitem [{\citenamefont {Karthika}\ and\ \citenamefont
  {Nagar}(2020)}]{Karthika_2020-reset}%
  \BibitemOpen
  \bibfield  {author} {\bibinfo {author} {\bibfnamefont {S.}~\bibnamefont
  {Karthika}}\ and\ \bibinfo {author} {\bibfnamefont {A.}~\bibnamefont
  {Nagar}},\ }\bibfield  {title} {\bibinfo {title} {Totally asymmetric simple
  exclusion process with resetting},\ }\href
  {https://doi.org/10.1088/1751-8121/ab6aef} {\bibfield  {journal} {\bibinfo
  {journal} {Journal of Physics A: Mathematical and Theoretical}\ }\textbf
  {\bibinfo {volume} {53}},\ \bibinfo {pages} {115003} (\bibinfo {year}
  {2020})}\BibitemShut {NoStop}%
\bibitem [{\citenamefont {Mishra}\ and\ \citenamefont
  {Basu}(2023)}]{Mishra_2023-reset}%
  \BibitemOpen
  \bibfield  {author} {\bibinfo {author} {\bibfnamefont {S.}~\bibnamefont
  {Mishra}}\ and\ \bibinfo {author} {\bibfnamefont {U.}~\bibnamefont {Basu}},\
  }\bibfield  {title} {\bibinfo {title} {Symmetric exclusion process under
  stochastic power-law resetting},\ }\href
  {https://doi.org/10.1088/1742-5468/accf06} {\bibfield  {journal} {\bibinfo
  {journal} {Journal of Statistical Mechanics: Theory and Experiment}\ }\textbf
  {\bibinfo {volume} {2023}},\ \bibinfo {pages} {053202} (\bibinfo {year}
  {2023})}\BibitemShut {NoStop}%
\bibitem [{\citenamefont {Magoni}\ \emph {et~al.}(2020)\citenamefont {Magoni},
  \citenamefont {Majumdar},\ and\ \citenamefont
  {Schehr}}]{PhysRevResearch.2.033182}%
  \BibitemOpen
  \bibfield  {author} {\bibinfo {author} {\bibfnamefont {M.}~\bibnamefont
  {Magoni}}, \bibinfo {author} {\bibfnamefont {S.~N.}\ \bibnamefont
  {Majumdar}},\ and\ \bibinfo {author} {\bibfnamefont {G.}~\bibnamefont
  {Schehr}},\ }\bibfield  {title} {\bibinfo {title} {Ising model with
  stochastic resetting},\ }\href
  {https://doi.org/10.1103/PhysRevResearch.2.033182} {\bibfield  {journal}
  {\bibinfo  {journal} {Phys. Rev. Res.}\ }\textbf {\bibinfo {volume} {2}},\
  \bibinfo {pages} {033182} (\bibinfo {year} {2020})}\BibitemShut {NoStop}%
\bibitem [{\citenamefont {Gupta}\ \emph {et~al.}(2014)\citenamefont {Gupta},
  \citenamefont {Majumdar},\ and\ \citenamefont
  {Schehr}}]{PhysRevLett.112.220601}%
  \BibitemOpen
  \bibfield  {author} {\bibinfo {author} {\bibfnamefont {S.}~\bibnamefont
  {Gupta}}, \bibinfo {author} {\bibfnamefont {S.~N.}\ \bibnamefont
  {Majumdar}},\ and\ \bibinfo {author} {\bibfnamefont {G.}~\bibnamefont
  {Schehr}},\ }\bibfield  {title} {\bibinfo {title} {Fluctuating interfaces
  subject to stochastic resetting},\ }\href
  {https://doi.org/10.1103/PhysRevLett.112.220601} {\bibfield  {journal}
  {\bibinfo  {journal} {Phys. Rev. Lett.}\ }\textbf {\bibinfo {volume} {112}},\
  \bibinfo {pages} {220601} (\bibinfo {year} {2014})}\BibitemShut {NoStop}%
\bibitem [{\citenamefont {Gupta}\ and\ \citenamefont
  {Nagar}(2016)}]{Gupta_2016_reset-FI}%
  \BibitemOpen
  \bibfield  {author} {\bibinfo {author} {\bibfnamefont {S.}~\bibnamefont
  {Gupta}}\ and\ \bibinfo {author} {\bibfnamefont {A.}~\bibnamefont {Nagar}},\
  }\bibfield  {title} {\bibinfo {title} {Resetting of fluctuating interfaces at
  power-law times},\ }\href {https://doi.org/10.1088/1751-8113/49/44/445001}
  {\bibfield  {journal} {\bibinfo  {journal} {Journal of Physics A:
  Mathematical and Theoretical}\ }\textbf {\bibinfo {volume} {49}},\ \bibinfo
  {pages} {445001} (\bibinfo {year} {2016})}\BibitemShut {NoStop}%
\bibitem [{\citenamefont {Mercado-Vásquez}\ and\ \citenamefont
  {Boyer}(2018)}]{Mercado-Vsquez_2018_PP}%
  \BibitemOpen
  \bibfield  {author} {\bibinfo {author} {\bibfnamefont {G.}~\bibnamefont
  {Mercado-Vásquez}}\ and\ \bibinfo {author} {\bibfnamefont {D.}~\bibnamefont
  {Boyer}},\ }\bibfield  {title} {\bibinfo {title} {Lotka–volterra systems
  with stochastic resetting},\ }\href
  {https://doi.org/10.1088/1751-8121/aadbc0} {\bibfield  {journal} {\bibinfo
  {journal} {Journal of Physics A: Mathematical and Theoretical}\ }\textbf
  {\bibinfo {volume} {51}},\ \bibinfo {pages} {405601} (\bibinfo {year}
  {2018})}\BibitemShut {NoStop}%
\bibitem [{\citenamefont {Evans}\ \emph {et~al.}(2022)\citenamefont {Evans},
  \citenamefont {Majumdar},\ and\ \citenamefont {Schehr}}]{Evans_2022-PP}%
  \BibitemOpen
  \bibfield  {author} {\bibinfo {author} {\bibfnamefont {M.~R.}\ \bibnamefont
  {Evans}}, \bibinfo {author} {\bibfnamefont {S.~N.}\ \bibnamefont
  {Majumdar}},\ and\ \bibinfo {author} {\bibfnamefont {G.}~\bibnamefont
  {Schehr}},\ }\bibfield  {title} {\bibinfo {title} {An exactly solvable
  predator prey model with resetting},\ }\href
  {https://doi.org/10.1088/1751-8121/ac7269} {\bibfield  {journal} {\bibinfo
  {journal} {Journal of Physics A: Mathematical and Theoretical}\ }\textbf
  {\bibinfo {volume} {55}},\ \bibinfo {pages} {274005} (\bibinfo {year}
  {2022})}\BibitemShut {NoStop}%
\bibitem [{\citenamefont {Singh}\ and\ \citenamefont
  {Singh}(2022)}]{PhysRevE.106.064118}%
  \BibitemOpen
  \bibfield  {author} {\bibinfo {author} {\bibfnamefont {R.~K.}\ \bibnamefont
  {Singh}}\ and\ \bibinfo {author} {\bibfnamefont {S.}~\bibnamefont {Singh}},\
  }\bibfield  {title} {\bibinfo {title} {Capture of a diffusing lamb by a
  diffusing lion when both return home},\ }\href
  {https://doi.org/10.1103/PhysRevE.106.064118} {\bibfield  {journal} {\bibinfo
   {journal} {Phys. Rev. E}\ }\textbf {\bibinfo {volume} {106}},\ \bibinfo
  {pages} {064118} (\bibinfo {year} {2022})}\BibitemShut {NoStop}%
\bibitem [{\citenamefont {Krapivsky}\ \emph {et~al.}(2022)\citenamefont
  {Krapivsky}, \citenamefont {Vilk},\ and\ \citenamefont
  {Meerson}}]{PhysRevE.106.034125}%
  \BibitemOpen
  \bibfield  {author} {\bibinfo {author} {\bibfnamefont {P.~L.}\ \bibnamefont
  {Krapivsky}}, \bibinfo {author} {\bibfnamefont {O.}~\bibnamefont {Vilk}},\
  and\ \bibinfo {author} {\bibfnamefont {B.}~\bibnamefont {Meerson}},\
  }\bibfield  {title} {\bibinfo {title} {Competition in a system of brownian
  particles: Encouraging achievers},\ }\href
  {https://doi.org/10.1103/PhysRevE.106.034125} {\bibfield  {journal} {\bibinfo
   {journal} {Phys. Rev. E}\ }\textbf {\bibinfo {volume} {106}},\ \bibinfo
  {pages} {034125} (\bibinfo {year} {2022})}\BibitemShut {NoStop}%
\bibitem [{\citenamefont {Miron}\ and\ \citenamefont
  {Reuveni}(2021)}]{PhysRevResearch.3.L012023}%
  \BibitemOpen
  \bibfield  {author} {\bibinfo {author} {\bibfnamefont {A.}~\bibnamefont
  {Miron}}\ and\ \bibinfo {author} {\bibfnamefont {S.}~\bibnamefont
  {Reuveni}},\ }\bibfield  {title} {\bibinfo {title} {Diffusion with local
  resetting and exclusion},\ }\href
  {https://doi.org/10.1103/PhysRevResearch.3.L012023} {\bibfield  {journal}
  {\bibinfo  {journal} {Phys. Rev. Res.}\ }\textbf {\bibinfo {volume} {3}},\
  \bibinfo {pages} {L012023} (\bibinfo {year} {2021})}\BibitemShut {NoStop}%
\bibitem [{\citenamefont {Pelizzola}\ \emph {et~al.}(2021)\citenamefont
  {Pelizzola}, \citenamefont {Pretti},\ and\ \citenamefont
  {Zamparo}}]{Pelizzola_2021-lr}%
  \BibitemOpen
  \bibfield  {author} {\bibinfo {author} {\bibfnamefont {A.}~\bibnamefont
  {Pelizzola}}, \bibinfo {author} {\bibfnamefont {M.}~\bibnamefont {Pretti}},\
  and\ \bibinfo {author} {\bibfnamefont {M.}~\bibnamefont {Zamparo}},\
  }\bibfield  {title} {\bibinfo {title} {Simple exclusion processes with local
  resetting},\ }\href {https://doi.org/10.1209/0295-5075/133/60003} {\bibfield
  {journal} {\bibinfo  {journal} {Europhysics Letters}\ }\textbf {\bibinfo
  {volume} {133}},\ \bibinfo {pages} {60003} (\bibinfo {year}
  {2021})}\BibitemShut {NoStop}%
\bibitem [{\citenamefont {Biroli}\ \emph {et~al.}(2023)\citenamefont {Biroli},
  \citenamefont {Larralde}, \citenamefont {Majumdar},\ and\ \citenamefont
  {Schehr}}]{PhysRevLett.130.207101}%
  \BibitemOpen
  \bibfield  {author} {\bibinfo {author} {\bibfnamefont {M.}~\bibnamefont
  {Biroli}}, \bibinfo {author} {\bibfnamefont {H.}~\bibnamefont {Larralde}},
  \bibinfo {author} {\bibfnamefont {S.~N.}\ \bibnamefont {Majumdar}},\ and\
  \bibinfo {author} {\bibfnamefont {G.}~\bibnamefont {Schehr}},\ }\bibfield
  {title} {\bibinfo {title} {Extreme statistics and spacing distribution in a
  brownian gas correlated by resetting},\ }\href
  {https://doi.org/10.1103/PhysRevLett.130.207101} {\bibfield  {journal}
  {\bibinfo  {journal} {Phys. Rev. Lett.}\ }\textbf {\bibinfo {volume} {130}},\
  \bibinfo {pages} {207101} (\bibinfo {year} {2023})}\BibitemShut {NoStop}%
\bibitem [{\citenamefont {Sch{\"u}tz}(2000)}]{Schutz}%
  \BibitemOpen
  \bibfield  {author} {\bibinfo {author} {\bibfnamefont {G.~M.}\ \bibnamefont
  {Sch{\"u}tz}},\ }\bibfield  {title} {\bibinfo {title} {Exact tracer diffusion
  coefficient in the asymmetric random average process},\ }\href
  {https://doi.org/10.1023/A:1018664117102} {\bibfield  {journal} {\bibinfo
  {journal} {Journal of Statistical Physics}\ }\textbf {\bibinfo {volume}
  {99}},\ \bibinfo {pages} {1045} (\bibinfo {year} {2000})}\BibitemShut
  {NoStop}%
\bibitem [{\citenamefont {Rajesh}\ and\ \citenamefont
  {Majumdar}(2001)}]{PhysRevE.64.036103}%
  \BibitemOpen
  \bibfield  {author} {\bibinfo {author} {\bibfnamefont {R.}~\bibnamefont
  {Rajesh}}\ and\ \bibinfo {author} {\bibfnamefont {S.~N.}\ \bibnamefont
  {Majumdar}},\ }\bibfield  {title} {\bibinfo {title} {Exact tagged particle
  correlations in the random average process},\ }\href
  {https://doi.org/10.1103/PhysRevE.64.036103} {\bibfield  {journal} {\bibinfo
  {journal} {Phys. Rev. E}\ }\textbf {\bibinfo {volume} {64}},\ \bibinfo
  {pages} {036103} (\bibinfo {year} {2001})}\BibitemShut {NoStop}%
\bibitem [{\citenamefont {Stojkoski}\ \emph
  {et~al.}(2022{\natexlab{b}})\citenamefont {Stojkoski}, \citenamefont
  {Jolakoski}, \citenamefont {Pal}, \citenamefont {Sandev}, \citenamefont
  {Kocarev},\ and\ \citenamefont {Metzler}}]{income-Pal}%
  \BibitemOpen
  \bibfield  {author} {\bibinfo {author} {\bibfnamefont {V.}~\bibnamefont
  {Stojkoski}}, \bibinfo {author} {\bibfnamefont {P.}~\bibnamefont
  {Jolakoski}}, \bibinfo {author} {\bibfnamefont {A.}~\bibnamefont {Pal}},
  \bibinfo {author} {\bibfnamefont {T.}~\bibnamefont {Sandev}}, \bibinfo
  {author} {\bibfnamefont {L.}~\bibnamefont {Kocarev}},\ and\ \bibinfo {author}
  {\bibfnamefont {R.}~\bibnamefont {Metzler}},\ }\bibfield  {title} {\bibinfo
  {title} {Income inequality and mobility in geometric brownian motion with
  stochastic resetting: theoretical results and empirical evidence of
  non-ergodicity},\ }\href {https://doi.org/10.1098/rsta.2021.0157} {\bibfield
  {journal} {\bibinfo  {journal} {Philosophical Transactions of the Royal
  Society A: Mathematical, Physical and Engineering Sciences}\ }\textbf
  {\bibinfo {volume} {380}},\ \bibinfo {pages} {20210157} (\bibinfo {year}
  {2022}{\natexlab{b}})}\BibitemShut {NoStop}%
\bibitem [{\citenamefont {Ferrari}\ and\ \citenamefont
  {Fontes}(1998)}]{10.1214/EJP.v3-28}%
  \BibitemOpen
  \bibfield  {author} {\bibinfo {author} {\bibfnamefont {P.}~\bibnamefont
  {Ferrari}}\ and\ \bibinfo {author} {\bibfnamefont {L.}~\bibnamefont
  {Fontes}},\ }\bibfield  {title} {\bibinfo {title} {{Fluctuations of a Surface
  Submitted to a Random Average Process}},\ }\href
  {https://doi.org/10.1214/EJP.v3-28} {\bibfield  {journal} {\bibinfo
  {journal} {Electronic Journal of Probability}\ }\textbf {\bibinfo {volume}
  {3}},\ \bibinfo {pages} {1 } (\bibinfo {year} {1998})}\BibitemShut {NoStop}%
\bibitem [{\citenamefont {Coppersmith}\ \emph {et~al.}(1996)\citenamefont
  {Coppersmith}, \citenamefont {Liu}, \citenamefont {Majumdar}, \citenamefont
  {Narayan},\ and\ \citenamefont {Witten}}]{PhysRevE.53.4673}%
  \BibitemOpen
  \bibfield  {author} {\bibinfo {author} {\bibfnamefont {S.~N.}\ \bibnamefont
  {Coppersmith}}, \bibinfo {author} {\bibfnamefont {C.~h.}\ \bibnamefont
  {Liu}}, \bibinfo {author} {\bibfnamefont {S.}~\bibnamefont {Majumdar}},
  \bibinfo {author} {\bibfnamefont {O.}~\bibnamefont {Narayan}},\ and\ \bibinfo
  {author} {\bibfnamefont {T.~A.}\ \bibnamefont {Witten}},\ }\bibfield  {title}
  {\bibinfo {title} {Model for force fluctuations in bead packs},\ }\href
  {https://doi.org/10.1103/PhysRevE.53.4673} {\bibfield  {journal} {\bibinfo
  {journal} {Phys. Rev. E}\ }\textbf {\bibinfo {volume} {53}},\ \bibinfo
  {pages} {4673} (\bibinfo {year} {1996})}\BibitemShut {NoStop}%
\bibitem [{\citenamefont {Rajesh}\ and\ \citenamefont
  {Majumdar}(2000)}]{Rajesh2000con}%
  \BibitemOpen
  \bibfield  {author} {\bibinfo {author} {\bibfnamefont {R.}~\bibnamefont
  {Rajesh}}\ and\ \bibinfo {author} {\bibfnamefont {S.~N.}\ \bibnamefont
  {Majumdar}},\ }\bibfield  {title} {\bibinfo {title} {Conserved mass models
  and particle systems in one dimension},\ }\href
  {https://doi.org/10.1023/A:1018651714376} {\bibfield  {journal} {\bibinfo
  {journal} {Journal of Statistical Physics}\ }\textbf {\bibinfo {volume}
  {99}},\ \bibinfo {pages} {943–965} (\bibinfo {year} {2000})}\BibitemShut
  {NoStop}%
\bibitem [{\citenamefont {Krug}\ and\ \citenamefont
  {Garcia}(2000)}]{Krug000con}%
  \BibitemOpen
  \bibfield  {author} {\bibinfo {author} {\bibfnamefont {J.}~\bibnamefont
  {Krug}}\ and\ \bibinfo {author} {\bibfnamefont {J.}~\bibnamefont {Garcia}},\
  }\bibfield  {title} {\bibinfo {title} {Asymmetric particle systems on r},\
  }\href {https://doi.org/10.1023/A:1018688421856} {\bibfield  {journal}
  {\bibinfo  {journal} {Journal of Statistical Physics}\ }\textbf {\bibinfo
  {volume} {99}},\ \bibinfo {pages} {31–55} (\bibinfo {year}
  {2000})}\BibitemShut {NoStop}%
\bibitem [{\citenamefont {Ispolatov}\ \emph {et~al.}(1998)\citenamefont
  {Ispolatov}, \citenamefont {Krapivsky},\ and\ \citenamefont
  {Redner}}]{Ispolatov1998}%
  \BibitemOpen
  \bibfield  {author} {\bibinfo {author} {\bibfnamefont {S.}~\bibnamefont
  {Ispolatov}}, \bibinfo {author} {\bibfnamefont {P.~L.}\ \bibnamefont
  {Krapivsky}},\ and\ \bibinfo {author} {\bibfnamefont {S.}~\bibnamefont
  {Redner}},\ }\bibfield  {title} {\bibinfo {title} {Wealth distributions in
  asset exchange models},\ }\href {https://doi.org/10.1007/s100510050249}
  {\bibfield  {journal} {\bibinfo  {journal} {The European Physical Journal B -
  Condensed Matter and Complex Systems}\ }\textbf {\bibinfo {volume} {2}},\
  \bibinfo {pages} {267} (\bibinfo {year} {1998})}\BibitemShut {NoStop}%
\bibitem [{\citenamefont {Aldous}\ and\ \citenamefont
  {Diaconis}(1995)}]{Hammersely1995}%
  \BibitemOpen
  \bibfield  {author} {\bibinfo {author} {\bibfnamefont {D.}~\bibnamefont
  {Aldous}}\ and\ \bibinfo {author} {\bibfnamefont {P.}~\bibnamefont
  {Diaconis}},\ }\bibfield  {title} {\bibinfo {title} {Hammersley's interacting
  particle process and longest increasing subsequences},\ }\href
  {https://doi.org/10.1007/BF01204214} {\bibfield  {journal} {\bibinfo
  {journal} {Probability Theory and Related Fields}\ }\textbf {\bibinfo
  {volume} {103}},\ \bibinfo {pages} {199} (\bibinfo {year}
  {1995})}\BibitemShut {NoStop}%
\bibitem [{\citenamefont {Cividini}\ \emph
  {et~al.}(2016{\natexlab{a}})\citenamefont {Cividini}, \citenamefont {Kundu},
  \citenamefont {Majumdar},\ and\ \citenamefont {Mukamel}}]{Cividini_2016_RAP}%
  \BibitemOpen
  \bibfield  {author} {\bibinfo {author} {\bibfnamefont {J.}~\bibnamefont
  {Cividini}}, \bibinfo {author} {\bibfnamefont {A.}~\bibnamefont {Kundu}},
  \bibinfo {author} {\bibfnamefont {S.~N.}\ \bibnamefont {Majumdar}},\ and\
  \bibinfo {author} {\bibfnamefont {D.}~\bibnamefont {Mukamel}},\ }\bibfield
  {title} {\bibinfo {title} {Exact gap statistics for the random average
  process on a ring with a tracer},\ }\href
  {https://doi.org/10.1088/1751-8113/49/8/085002} {\bibfield  {journal}
  {\bibinfo  {journal} {Journal of Physics A: Mathematical and Theoretical}\
  }\textbf {\bibinfo {volume} {49}},\ \bibinfo {pages} {085002} (\bibinfo
  {year} {2016}{\natexlab{a}})}\BibitemShut {NoStop}%
\bibitem [{\citenamefont {Kundu}\ and\ \citenamefont
  {Cividini}(2016)}]{Kundu_2016_RAP}%
  \BibitemOpen
  \bibfield  {author} {\bibinfo {author} {\bibfnamefont {A.}~\bibnamefont
  {Kundu}}\ and\ \bibinfo {author} {\bibfnamefont {J.}~\bibnamefont
  {Cividini}},\ }\bibfield  {title} {\bibinfo {title} {Exact correlations in a
  single-file system with a driven tracer},\ }\href
  {https://doi.org/10.1209/0295-5075/115/54003} {\bibfield  {journal} {\bibinfo
   {journal} {Europhysics Letters}\ }\textbf {\bibinfo {volume} {115}},\
  \bibinfo {pages} {54003} (\bibinfo {year} {2016})}\BibitemShut {NoStop}%
\bibitem [{\citenamefont {Cividini}\ \emph
  {et~al.}(2016{\natexlab{b}})\citenamefont {Cividini}, \citenamefont {Kundu},
  \citenamefont {Majumdar},\ and\ \citenamefont
  {Mukamel}}]{Cividini_2016_RAPP}%
  \BibitemOpen
  \bibfield  {author} {\bibinfo {author} {\bibfnamefont {J.}~\bibnamefont
  {Cividini}}, \bibinfo {author} {\bibfnamefont {A.}~\bibnamefont {Kundu}},
  \bibinfo {author} {\bibfnamefont {S.~N.}\ \bibnamefont {Majumdar}},\ and\
  \bibinfo {author} {\bibfnamefont {D.}~\bibnamefont {Mukamel}},\ }\bibfield
  {title} {\bibinfo {title} {Correlation and fluctuation in a random average
  process on an infinite line with a driven tracer},\ }\href
  {https://doi.org/10.1088/1742-5468/2016/05/053212} {\bibfield  {journal}
  {\bibinfo  {journal} {Journal of Statistical Mechanics: Theory and
  Experiment}\ }\textbf {\bibinfo {volume} {2016}},\ \bibinfo {pages} {053212}
  (\bibinfo {year} {2016}{\natexlab{b}})}\BibitemShut {NoStop}%
\bibitem [{\citenamefont {Cividini}\ and\ \citenamefont
  {Kundu}(2017)}]{Cividini_2017_RAP}%
  \BibitemOpen
  \bibfield  {author} {\bibinfo {author} {\bibfnamefont {J.}~\bibnamefont
  {Cividini}}\ and\ \bibinfo {author} {\bibfnamefont {A.}~\bibnamefont
  {Kundu}},\ }\bibfield  {title} {\bibinfo {title} {Tagged particle in
  single-file diffusion with arbitrary initial conditions},\ }\href
  {https://doi.org/10.1088/1742-5468/aa75de} {\bibfield  {journal} {\bibinfo
  {journal} {Journal of Statistical Mechanics: Theory and Experiment}\ }\textbf
  {\bibinfo {volume} {2017}},\ \bibinfo {pages} {083203} (\bibinfo {year}
  {2017})}\BibitemShut {NoStop}%
\bibitem [{\citenamefont {Bateman}\ and\ \citenamefont
  {Project}(2023)}]{bateman_2023_mhd23-e0z22}%
  \BibitemOpen
  \bibfield  {author} {\bibinfo {author} {\bibfnamefont {H.}~\bibnamefont
  {Bateman}}\ and\ \bibinfo {author} {\bibfnamefont {B.~M.}\ \bibnamefont
  {Project}},\ }\href@noop {} {\emph {\bibinfo {title} {Tables of Integral
  Transforms}}}\ (\bibinfo  {publisher} {McGraw-Hill Book Company},\ \bibinfo
  {year} {2023})\BibitemShut {NoStop}%
\bibitem [{\citenamefont {Singh}\ and\ \citenamefont
  {Kundu}(2021)}]{Singh_2021_cross}%
  \BibitemOpen
  \bibfield  {author} {\bibinfo {author} {\bibfnamefont {P.}~\bibnamefont
  {Singh}}\ and\ \bibinfo {author} {\bibfnamefont {A.}~\bibnamefont {Kundu}},\
  }\bibfield  {title} {\bibinfo {title} {Crossover behaviours exhibited by
  fluctuations and correlations in a chain of active particles},\ }\href
  {https://doi.org/10.1088/1751-8121/ac0a9f} {\bibfield  {journal} {\bibinfo
  {journal} {Journal of Physics A: Mathematical and Theoretical}\ }\textbf
  {\bibinfo {volume} {54}},\ \bibinfo {pages} {305001} (\bibinfo {year}
  {2021})}\BibitemShut {NoStop}%
\bibitem [{\citenamefont {Put}\ \emph {et~al.}(2019)\citenamefont {Put},
  \citenamefont {Berx},\ and\ \citenamefont {Vanderzande}}]{Put_2019}%
  \BibitemOpen
  \bibfield  {author} {\bibinfo {author} {\bibfnamefont {S.}~\bibnamefont
  {Put}}, \bibinfo {author} {\bibfnamefont {J.}~\bibnamefont {Berx}},\ and\
  \bibinfo {author} {\bibfnamefont {C.}~\bibnamefont {Vanderzande}},\
  }\bibfield  {title} {\bibinfo {title} {Non-gaussian anomalous dynamics in
  systems of interacting run-and-tumble particles},\ }\href
  {https://doi.org/10.1088/1742-5468/ab4e90} {\bibfield  {journal} {\bibinfo
  {journal} {Journal of Statistical Mechanics: Theory and Experiment}\ }\textbf
  {\bibinfo {volume} {2019}},\ \bibinfo {pages} {123205} (\bibinfo {year}
  {2019})}\BibitemShut {NoStop}%
\bibitem [{\citenamefont {Santra}\ \emph {et~al.}(2023)\citenamefont {Santra},
  \citenamefont {Singh},\ and\ \citenamefont {Kundu}}]{Santra2023RAP}%
  \BibitemOpen
  \bibfield  {author} {\bibinfo {author} {\bibfnamefont {S.}~\bibnamefont
  {Santra}}, \bibinfo {author} {\bibfnamefont {P.}~\bibnamefont {Singh}},\ and\
  \bibinfo {author} {\bibfnamefont {A.}~\bibnamefont {Kundu}},\ }\href@noop {}
  {\bibinfo {title} {Tracer dynamics in active random average process}},\
  \bibinfo {howpublished} {\url{https://arxiv.org/abs/2307.09908}} (\bibinfo
  {year} {2023}),\ \bibinfo {note} {arXiv:2307.09908}\BibitemShut {NoStop}%
\end{thebibliography}%
%\end{thebibliography}

\end{document}